\documentclass[journal=jctcce,manuscript=article,layout=traditional]{achemso}
\usepackage{graphicx}
\usepackage{tabularx}
\usepackage{dcolumn}
\usepackage{tensor}
\usepackage{color}
\usepackage{placeins}
\usepackage{bm}
\usepackage{amsmath}
\usepackage{amsfonts}
\usepackage{amssymb}
\usepackage{float}
\providecommand{\U}[1]{\protect\rule{.1in}{.1in}}

\graphicspath{{"d:/Group Vanicek/Desktop/Azulene/figures/"}
{./figures/}{C:/Users/Jiri/Dropbox/Papers/Chemistry_papers/2019/AzulenePaper/Azulene/figures/}}

\title{Semiclassical Approach to Photophysics Beyond Kasha's Rule and Vibronic
Spectroscopy Beyond the Condon Approximation. The Case of Azulene}
\author{Antonio Prlj}
\altaffiliation{These authors contributed equally.}
\author{Tomislav Begu\v{s}i\'{c}}
\altaffiliation{These authors contributed equally.}
\author{Zhan Tong Zhang}
\affiliation{Laboratory of Theoretical Physical Chemistry, Institut des Sciences et
Ing\'enierie Chimiques, Ecole Polytechnique F\'ed\'erale de Lausanne (EPFL),
CH-1015, Lausanne, Switzerland}
\author{George Cameron Fish}
\affiliation{Photochemical Dynamics Group, Institut des Sciences et Ing\'enierie Chimiques,
Ecole Polytechnique F\'ed\'erale de Lausanne (EPFL), CH-1015, Lausanne, Switzerland}
\author{Marius Wehrle}
\author{Tom\'{a}\v{s} Zimmermann}
\author{Seonghoon Choi}
\author{Julien Roulet}
\affiliation{Laboratory of Theoretical Physical Chemistry, Institut des Sciences et
Ing\'enierie Chimiques, Ecole Polytechnique F\'ed\'erale de Lausanne (EPFL),
CH-1015, Lausanne, Switzerland}
\author{Jacques-Edouard Moser}
\email{je.moser@epfl.ch}
\affiliation{Photochemical Dynamics Group, Institut des Sciences et Ing\'enierie Chimiques,
Ecole Polytechnique F\'ed\'erale de Lausanne (EPFL), CH-1015, Lausanne, Switzerland}
\author{Ji\v{r}\'i Van\'i\v{c}ek}
\email{jiri.vanicek@epfl.ch}
\affiliation{Laboratory of Theoretical Physical Chemistry, Institut des Sciences et
Ing\'enierie Chimiques, Ecole Polytechnique F\'ed\'erale de Lausanne (EPFL),
CH-1015, Lausanne, Switzerland}
\date{\today}
\begin{tocentry}
\includegraphics[height=3.65cm,keepaspectratio]{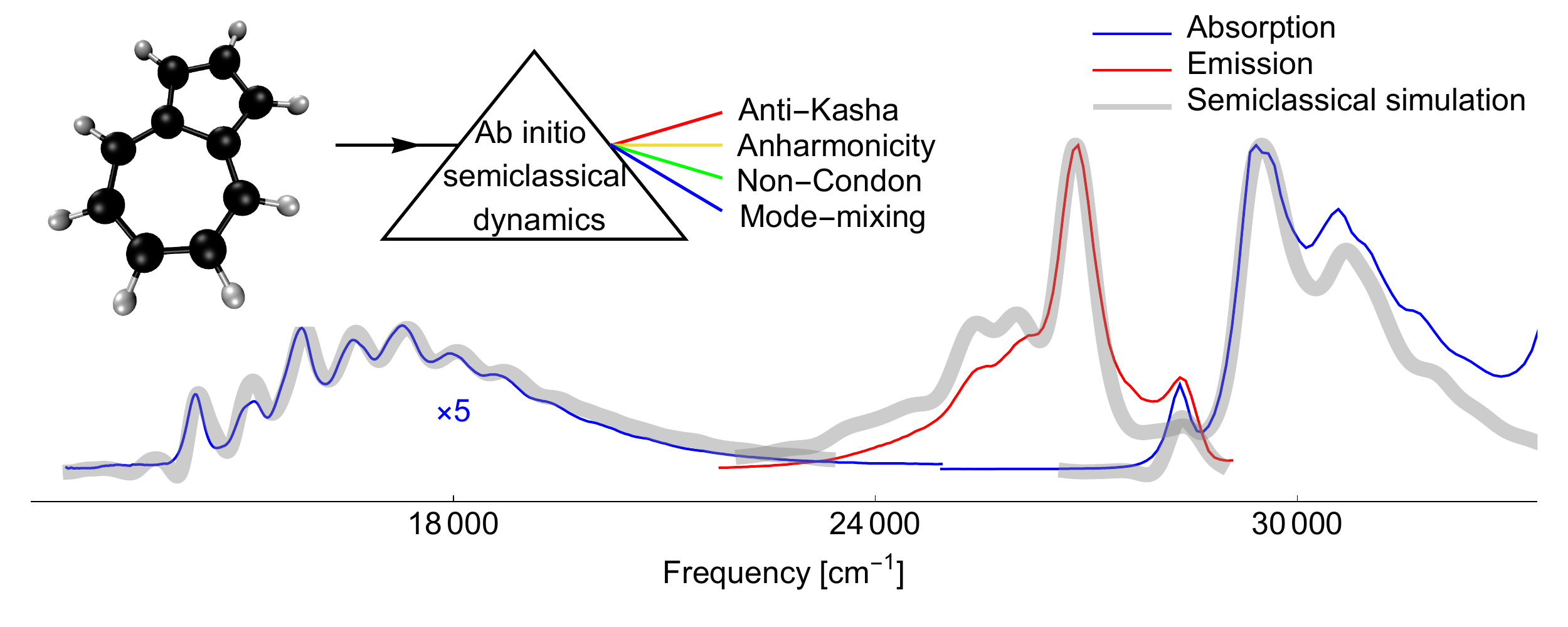}
\end{tocentry}

\begin{document}
\immediate\write18{texcount.pl Azulene_v14.tex}	
\begin{abstract}
Azulene is a prototypical molecule with an anomalous fluorescence from the
second excited electronic state, thus violating Kasha's rule, and with an
emission spectrum that cannot be understood within the Condon approximation.
To better understand photophysics and spectroscopy of azulene and other
non-conventional molecules, we develop a systematic, general, and efficient
computational approach combining semiclassical dynamics of nuclei with ab initio
electronic structure. First, to analyze the nonadiabatic effects, we complement the standard population dynamics by a
rigorous measure of adiabaticity, estimated with the multiple-surface
dephasing representation. Second, we propose a new semiclassical method for
simulating non-Condon spectra, which combines the
extended thawed Gaussian approximation with the efficient
single-Hessian approach. S$_{1} \leftarrow$ S$_0$ and S$_{2} \leftarrow$ S$_0$ absorption and S$_{2} \rightarrow$ S$_0$ emission
spectra of azulene, recorded in a new set of experiments, agree very well with
our calculations. We find that accuracy of the evaluated spectra requires
the treatment of anharmonicity, Herzberg--Teller, and mode-mixing effects.
\end{abstract}

\section{\label{sec:intro}Introduction}
Azulene molecule is an archetypal system violating
Kasha's rule,\cite{Beer_Longuet_Higgins:1955,Viswanath_Kasha:1956} according
to which \textquotedblleft polyatomic molecular entities luminesce with
appreciable yield only from the lowest excited state of a given
multiplicity.\textquotedblright\cite{book_GoldBook} As a result, azulene has
attracted significant
experimental\cite{Ippen_Woerner1977,Amirav_Jortner:1984,Huppert_Rentzepis:1972,Diau_Zewail:1999,Rentzepis:1969,Foggi_Salvi:2003,Vosskotter_Weinkauf:2015,Gillispie_Lim:1978,Klemp_Nickel:1983}
and
theoretical\cite{Foggi_Salvi:2003,Vosskotter_Weinkauf:2015,Bearpark_Vreven:1996,Klein_Bernardi:1998,Gustav_Storch:1990,Negri_Zgierski:1993,Amatatsu_Komura:2006,Murakami_Nakamura:2004}
attention over the decades. More recently, rigorous experimental and theoretical approaches proved useful in identifying, but also refuting, the violation of Kasha's rule in other molecular systems.\cite{DelValle_Catalan:2019,Paul_Chakrabarti:2019,Rohrs_Escudero:2019,Shafikov_Czerwieniec:2019, Zhou_Han:2019}

\begin{figure}
\includegraphics[width=0.48\textwidth]{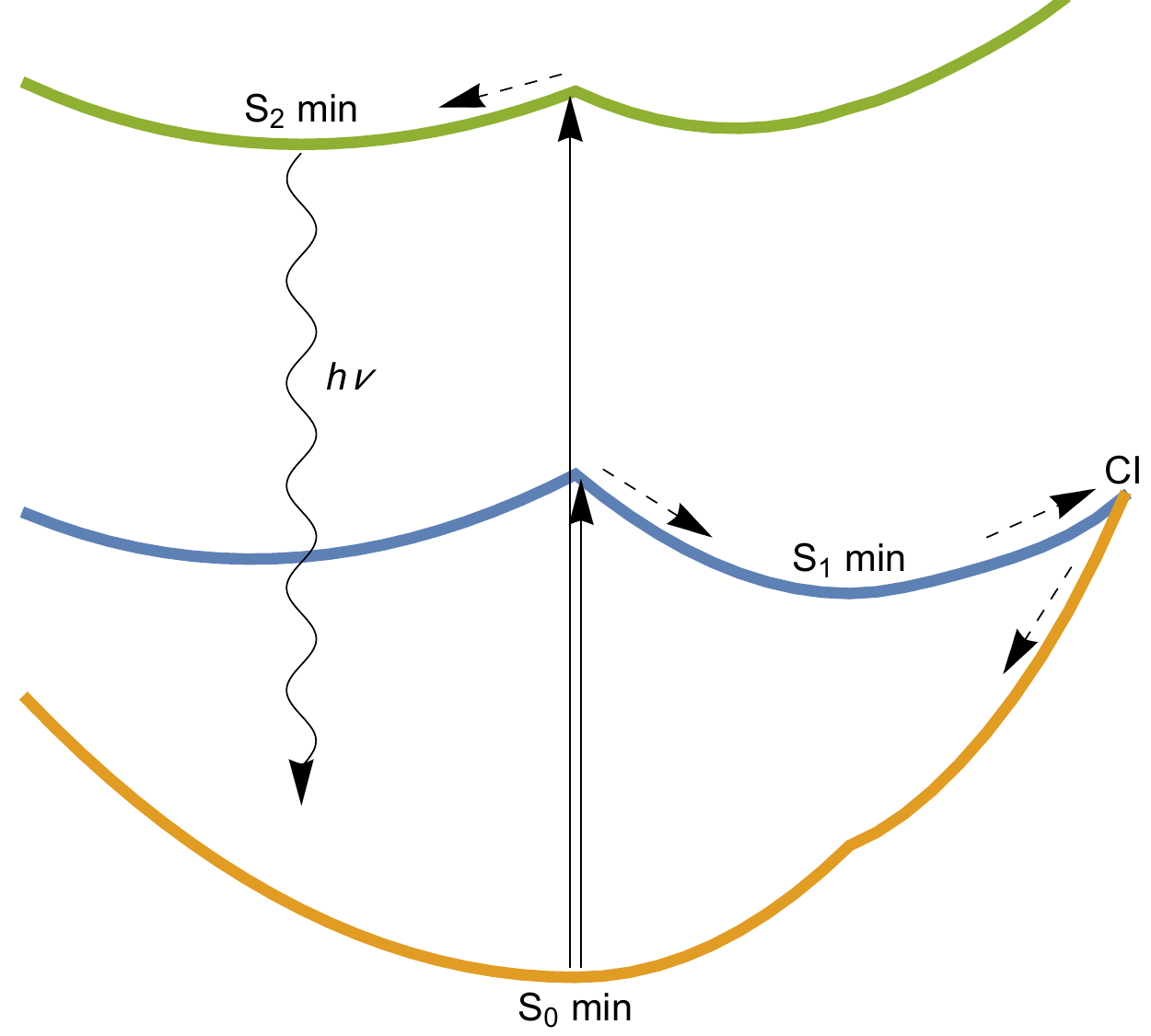}
\caption{\label{fig:Scheme1}Schematic representation of azulene photophysics involving S$_1$ (right) and S$_2$ state (left). Absorption, emission, and nuclear dynamics are represented by full, curved, and dashed arrows, respectively. Potential energy surface cuts are based on CASSCF electronic structure (see Supporting Information for details). Note that the calculations of adiabaticity, population dynamics, and spectra presented in the results section are based on full-dimensional ab initio potential energy surfaces and not on this schematic representation.}
\end{figure}

Spectroscopic and photophysical studies tried to explain why observed
fluorescence in azulene occurs from the second (S$_{2}$) instead of the first (S$_{1}$) excited singlet state. The measured
lifetimes of the S$_{1}$ state of azulene range from $\sim2$ ps in
solution\cite{Ippen_Woerner1977} to $\sim1$ ps in the gas
phase,\cite{Amirav_Jortner:1984,Diau_Zewail:1999} indicating that
radiationless decay is much faster than the time scale of emission itself.
Surface-hopping and Ehrenfest simulations by Robb \textit{et al.}%
\cite{Bearpark_Vreven:1996,Klein_Bernardi:1998} ascribed the ultrafast decay
to the energetically low-lying conical intersection (see Fig.~\ref{fig:Scheme1}) between the S$_{1}$ state
and the ground electronic state, S$_{0}$, although the estimated S$_{1}$
lifetime ($\sim10$ fs) was significantly smaller than the experimental one.
Apart from the S$_{1}$ fluorescence quenching, which is ubiquitous in a wide
range of small and medium-sized organic molecules, anomalous behavior of
azulene shows itself in the characteristic fluorescence from the S$_{2}$
state. Hindered S$_{2}$ $\rightarrow$ S$_{1}$ internal conversion is
attributed to the wide interstate gap (see Fig.~\ref{fig:Scheme1}) and, more precisely, to the weak
nonadiabatic coupling (NAC), giving rise to the moderate, yet distinctive,
S$_{2}$ emission. It was estimated that fluorescence quantum yield $\Phi_{f}%
$(S$_{2}$) of the second excited state outcompetes $\Phi_{f}$(S$_{1}$) by four
orders of magnitude, while the nonradiative internal conversion constant
$k_{\text{IC}}^{2\rightarrow1}$ is 100 times smaller than $k_{\text{IC}%
}^{1\rightarrow0}$.\cite{Gustav_Storch:1990} The S$_{1} \leftarrow$ S$_{0}$ absorption spectrum
was correctly reproduced by Franck-Condon
simulations,\cite{Dierksen_Grimme:2004,Niu_Shuai:2010} assuming the validity
of Condon approximation,\cite{Condon:1928} which neglects the dependence of
the transition dipole moment on nuclear coordinates. The most comprehensive
study of importance of non-Condon effects in azulene was the early work of
Gustav and Storch,\cite{Gustav_Storch:1990} who showed that S$_{1}$ absorption
and emission have dominant Condon contributions, while S$_{2} \rightarrow$ S$_0$ emission has
important Herzberg-Teller effects. S$_{2}$ $\leftarrow$ S$_{0}$
absorption was not considered.

Motivated by its unusual photophysics and spectroscopy, we use azulene as a
test case for introducing an efficient computational approach for studying
various Kasha violating (or Kasha-obeying) systems.
Such theoretical tool seems necessary since experimental verification of anti-Kasha behavior may be quite
challenging, as demostrated by recent reports.\cite{Shafikov_Czerwieniec:2019,Zhou_Han:2019}
The proposed methodology consists of two steps: (i) To analyze the influence of NACs on the
excited-state dynamics, we evaluate the \textquotedblleft
adiabaticity\textquotedblright\ with a rigorous measure that is approximately
evaluated semiclassically with the multiple-surface dephasing representation
(MSDR).\cite{Zimmermann_Vanicek:2012,Zimmermann_Vanicek:2012a} Because it can
detect more subtle nonadiabatic effects, the adiabaticity goes beyond the
standard analysis based on population
dynamics.\cite{Martens_Fang:1997,Muller_Stock:1997,Worth_Robb:2003,Tapavicza_Casida:2008,Richter_Gonzalez:2011,Curchod_Rothlisberger:2011,Belyaev_Triglia:2014,Richings_Lasorne:2015,Curchod_Tavernelli:2018,Glover_Martinez:2018}
(ii) Building on the joint analysis of adiabaticity and population dynamics,
we introduce a new method for computing vibrationally resolved electronic
spectra by combining the single-Hessian\cite{Begusic_Vanicek:2019} and
extended\cite{Lee_Heller:1982,Patoz_Vanicek:2018, Begusic_Vanicek:2018} thawed
Gaussian approximations. The new methodology, augmented with on-the-fly
\textit{ab initio} electronic structure calculations, is applied to study
nonadiabatic, non-Condon, anharmonicity, and mode-mixing effects in the first
two excited states of azulene.

\section{\label{sec:theory}Theory}

\subsection{\label{subsec:MSDR}Measuring adiabaticity with multiple-surface dephasing representation}

A natural way to estimate the effect of NACs on the molecular quantum dynamics
launched from a certain electronic state is to analyze the subsequent
population dynamics. In higher dimensions, the time dependence of populations
is most often approximated with mixed quantum-classical methods, in which the
molecular wavefunction $\bm{\Psi}$ is replaced with an ensemble of $N$
trajectories, each of which is characterized by the classical nuclear position
($q$) and momentum ($p$), propagated with Hamilton's equations of motion,
\begin{equation}
\dot{q}_{j}(t)=\frac{\partial H^{(j)}(q_{j},p_{j})}{\partial p_{j}}, \quad
\dot{p}_{j}(t)=-\frac{\partial H^{(j)}(q_{j},p_{j})}{\partial q_{j}},\label{eq:MQC_qp}
\end{equation}
and by the electronic wavefunction $\mathbf{c}$, propagated with the
time-dependent Schr\"{o}dinger equation
\begin{equation}
i\hbar\mathbf{\dot{c}}_{j}(t)=\mathbf{H}(q_{j}(t),p_{j}(t))\mathbf{c}_{j}(t).\label{eq:MQC_c}
\end{equation}
Here $j=1,\ldots,N$ is the index of the trajectory, $H^{(j)}$ is a
method-dependent approximate Hamiltonian associated with the $j$th trajectory,
and $\mathbf{H}$ denotes the fully coupled molecular Hamiltonian expressed in
the basis of the $S$ considered electronic states. In general, the bold face
denotes either the $S$-component vectors (e.g., $\mathbf{c}$) or $S\times S$ matrices (e.g. $\mathbf{H}$) acting on the Hilbert space
spanned by the $S$ electronic states. While Ehrenfest dynamics evolves $q$ and
$p$ with the locally mean-field Hamiltonian $H^{(j)}_{\text{Ehr}}:=\langle
\mathbf{H}(q_{j}(t),p_{j}(t))\rangle_{\mathbf{c}_{j}(t)}$, where $\left\langle
\mathbf{A}\right\rangle _{\mathbf{c}}:=\mathbf{c}^{\dag}\mathbf{Ac}$ denotes
the expectation value of electronic operator $\mathbf{A}$ in the state
$\mathbf{c}$, both Born--Oppenheimer and surface hopping\cite{Tully:1990} algorithms employ the
(diagonal)\ Born-Oppenheimer Hamiltonian $H^{(j)}_{\text{BO}} \equiv H^{(j)}_{\text{SH}}:=\mathbf{H}%
_{n_{j}(t)n_{j}(t)}^{\text{BO}}(q_{j}(t),p_{j}(t))$, where $n_{j}%
(t)\in\{1,\ldots,S\}$ is the index of the adiabatic potential energy surface
on which the trajectory runs. In addition, in surface hopping, a stochastic
algorithm\cite{Tully:1990} is used to switch (or keep fixed) the current
surface $n_{j}(t)$ according to the current value of $\mathbf{c}_{j}%
(t)$, and a so-called \textquotedblleft decoherence
correction\textquotedblright\ \cite{Crespo-Otero_Barbatti:2018} is frequently added to improve the accuracy and consistency between the populations
obtained from the electronic wavefunctions $\mathbf{c}_{j}$ (\textquotedblleft
quantum populations\textquotedblright) and from the histogram of $n_{j}$
(\textquotedblleft classical populations\textquotedblright).

However, the NACs may affect more than just the populations of different
electronic states. A more rigorous measure of the importance of NACs is,
therefore, the \textquotedblleft adiabaticity,\textquotedblright
\begin{equation}
A(t):=\left\vert a(t)\right\vert ^{2}, \label{eq:adiab}
\end{equation}
where 
\begin{equation}
a(t)=\langle\bm{\Psi}(t)|\bm{\Psi}^{\text{BO}}(t)\rangle \label{eq:adiab_amp}
\end{equation} 
is the overlap of molecular wavefunctions propagated
either exactly or within the Born-Oppenheimer
approximation.\cite{Zimmermann_Vanicek:2010,Zimmermann_Vanicek:2012,MacKenzie_Pineault:2012}
More precisely, $|\bm{\Psi}(t)\rangle=e^{-i{\mathbf{\hat{H}}}t/\hbar}%
|\bm{\Psi}(0)\rangle,$ where $\mathbf{\hat{H}}$ is the fully coupled nonadiabatic
molecular Hamiltonian and $|\bm{\Psi}^{\text{BO}}(t)\rangle=e^{-i{\mathbf{\hat{H}}%
}^{\text{BO}}t/\hbar}|\bm{\Psi}(0)\rangle$, where $\mathbf{\hat{H}}^{\text{BO}}$ is
the Born-Oppenheimer Hamiltonian, in which the NACs are neglected. The hat
$\hat{}$ denotes nuclear operators. Obviously, for two normalized wave
packets, the adiabaticity $A$ is a number between 0 and 1, where high
adiabaticity, $A(t)\approx1$, indicates that the Born-Oppenheimer
approximation at time $t$ is accurate, whereas low adiabaticity, $A(t)\ll1$,
suggests that nonadiabatic couplings are important and should be taken into
account in an accurate simulation.

Evaluating adiabaticity $A(t)$ exactly in higher-dimensional systems is a
formidable, if not impossible, task because it requires exact quantum
propagation. Fortunately, the semiclassical MSDR provides, in many situations,
a very good estimate of adiabaticity at the fraction of the cost of exact
quantum calculation.\cite{Zimmermann_Vanicek:2012,Zimmermann_Vanicek:2012a}
Moreover, this semiclassical estimate of adiabaticity amplitude $\langle
\bm{\Psi}(t)|\bm{\Psi}^{\text{BO}}(t)\rangle$ is, typically, much more accurate than the
semiclassical approximations to the wavefunctions $\bm{\Psi}(t)$ and
$\bm{\Psi}^{\text{BO}}(t)$ themselves. Within the MSDR, the adiabaticity amplitude $a$ is
approximated as
\begin{equation}
a_{\text{MSDR}}(t)=h^{-D}\operatorname*{Tr}{}_{e}\int dx\bm{\rho}_{W}%
^{\text{init}}(x)\mathcal{T}e^{i\int_{0}^{t}\Delta\mathbf{H}_{W}^{I}(x,t^{\prime
})dt^{\prime}/\hbar},\label{eq:MSDR}%
\end{equation}
where $D$ is the number of nuclear degrees of freedom, $\operatorname*{Tr}%
_{e}$ denotes the trace over electronic degrees of freedom (the $S$ electronic
states here), $x=(q,p)$ denotes the $2D$ nuclear phase space coordinates at time $t$, and $\mathcal{T}$
is the time ordering operator. In addition, $\bm{\rho}^{\text{init}}$ is a density
operator of the initial state, $\Delta\mathbf{\hat{H}}:={\mathbf{\hat{H}}%
}-{\mathbf{\hat{H}}}^{\text{BO}}$ is the difference between the exact and
Born-Oppenheimer Hamiltonians, superscript $I$ denotes the interaction picture,
and subscript $W$ indicates a partial Wigner
transform\cite{Zimmermann_Vanicek:2012} over nuclear degrees of freedom. In
the most common case of electronically pure
states,\cite{Zimmermann_Vanicek:2012} the MSDR of adiabaticity can be
evaluated simply as\cite{Zimmermann_Vanicek:2012}
\begin{equation}
a_{\text{MSDR}}(t)=\overline{\mathbf{c}(t)^{\dag}\mathbf{c}_{\text{BO}}%
(t)},\label{eq:MSDR_c}%
\end{equation}
where the overbar denotes an average over the ensemble of trajectories,
$\overline{A}:=N^{-1}\sum_{j=1}^{N}A_{j}$, while $\mathbf{c}_{\text{BO}}(t)$
is the electronic wavefunction propagated with Eq.~(\ref{eq:MQC_c}) in which
the full Hamiltonian $\mathbf{H}$ is replaced with $\mathbf{H}^{\text{BO}}$.
As for the nuclear trajectories $(q,p)$, they can be propagated with the
fewest-switches surface hopping, Ehrenfest, or Born-Oppenheimer dynamics.
Overall, the MSDR allows quantitative analysis of the importance of NACs (and
beyond\cite{Vanicek:2017}), adding little additional cost to the (classical)
nuclear dynamics itself, while approximately introducing nuclear quantum
effects.\cite{Zimmermann_Vanicek:2012}

\subsection{\label{subsec:spec}Vibrationally resolved electronic spectroscopy}

The usual time-dependent approach to one-photon
spectroscopy\cite{Heller:1981a} treats the light-matter interaction within the
first-order perturbation theory. While it is equivalent to the
time-independent Franck--Condon approach, the time-dependent approach unravels
the direct relationship between vibrationally resolved electronic spectra and
molecular wavepacket dynamics. In the zero-temperature limit, i.e., assuming
only the state $|1, g \rangle$, the ground ($g$) vibrational state of the
ground (1) electronic state, is populated before the interaction with the
electromagnetic field, the linear absorption cross-section can be evaluated
as\cite{Heller:1981a,book_Tannor:2007,Lami_Santoro:2004,Niu_Shuai:2010}
\begin{equation}
\sigma^{\text{abs}}(\vec{\epsilon}, \omega) = \frac{4 \pi\omega}{ \hbar c}
\text{Re} \int_{0}^{\infty} dt C (\vec{\epsilon}, t) e^{i (\omega+ \omega_{1,
g}) t}. \label{eq:sigma_abs}%
\end{equation}
Here
\begin{equation}
C (\vec{\epsilon}, t) = \langle\phi(0) | \phi(t) \rangle\label{eq:C_t}%
\end{equation}
is the wavepacket autocorrelation function for the initial nuclear wavepacket
$|\phi(0) \rangle= \hat{\mu} |1, g \rangle$ evolved with the excited-state
nuclear Hamiltonian $\hat{H}_{2}$, $\hat{\mu}$ is the transition dipole moment matrix element $\hat{\vec{\mu}%
}_{21}$ projected on the three-dimensional polarization unit vector
$\vec{\epsilon}$ of the electric field, i.e., $\hat{\mu} = \hat{\vec{\mu}%
}_{21} \cdot\vec{\epsilon}$, and $\hbar\omega_{1, g} = \langle1, g | \hat
{H}_{1} | 1, g \rangle$ is the zero point energy. Emission spectrum, expressed as the emission rate per unit frequency, is
computed similarly,\cite{Lami_Santoro:2004,Niu_Shuai:2010} as
\begin{equation}
\sigma^{\text{em}}(\vec{\epsilon}, \omega) = \frac{4 \omega^{3}}{ \pi\hbar
c^{3}} \text{Re} \int_{0}^{\infty} dt C (\vec{\epsilon}, t)^{*} e^{i (\omega-
\omega_{2, g}) t}, \label{eq:sigma_em}%
\end{equation}
where the autocorrelation function $C(\vec{\epsilon}, t)$ is still given by
Eq.~(\ref{eq:C_t}), but the initial state $|\phi(0) \rangle= \hat{\mu} |2, g
\rangle$, obtained by multiplying the ground ($g$) vibrational state of an
excited (2) electronic state by the transition dipole moment, is propagated on
the ground-state surface. Finally, the spectrum averaged over all molecular orientations is evaluated simply as\cite{Craig_Thirunamachandran:1984, Begusic_Vanicek:2018} 
$\sigma_{\text{av.}} (\omega) = (1/3) \sum_{i} \sigma(\vec{e}_{i}, \omega)$,
where $\vec{e}_{i}$ ($i = x, y, z$) denotes the unit vector along the $i$-axis.

Different methods exist for simulating vibrationally resolved absorption and emission spectra of polyatomic molecules. The most standard approach is based on constructing global harmonic models\cite{Santoro_Barone:2007,AvilaFerrer_Santoro:2012,Baiardi_Barone:2013,Santoro_Jacquemin:2016,Benkyi_Sundholm:2019,Tapavicza:2019} for the ground- and excited-state potential energy surfaces, which requires only a few \textit{ab initio} calculations. The main advantages of the harmonic approximation are the existence of analytical expressions for the autocorrelation functions and the straightforward incorporation of temperature effects at nearly no additional cost. However, the method neglects potentially significant anharmonicity effects.

In an earlier work in our
group,\cite{Vanicek:2017,Zimmermann_Vanicek:2014} we showed that the
semiclassical MSDR, after a small extension, could be used to approximate
vibronic spectra, including nonadiabatic effects, but missing high resolution features. In contrast, the
thawed Gaussian approximation,\cite{Heller:1975} is rather accurate at
reproducing moderately resolved vibronic spectra,\cite{Wehrle_Vanicek:2014,
Wehrle_Vanicek:2015, Patoz_Vanicek:2018, Begusic_Vanicek:2018} but cannot
account for the nonadiabatic effects. As a result, the thawed Gaussian
propagation is limited to systems in which the Born--Oppenheimer approximation
holds; in such systems, however, it consistently outperforms commonly used
global harmonic methods because it can partially account for the anharmonicity of the potential energy surface.

\subsection{\label{subsec:tga}Evaluating spectra beyond Condon and harmonic approximations with single-Hessian extended thawed Gaussian approximation}

The thawed Gaussian approximation propagates a Gaussian wavepacket
\begin{equation}
\psi(q, t) = \frac{1}{(\pi \hbar)^{D/4} \sqrt{\det Q_{t}}} \exp\left\{  \frac
{i}{\hbar} \left[  \frac{1}{2} (q-q_{t})^{T} \cdot P_{t} \cdot Q_{t}^{-1}
\cdot(q-q_{t}) + p_{t}^{T} \cdot(q-q_{t}) + S_{t} \right]
\right\}  , \label{eq:gwp}%
\end{equation}
here written using Hagedorn's
parametrization,\cite{Heller:1976a,Hagedorn:1980,Hagedorn:1998,Faou_Lubich:2009}
in an effective time-dependent potential given by the local harmonic
approximation
\begin{equation}
V_{\text{LHA}}(q, t) = V(q_{t}) + V^{\prime} (q_{t})^{T} \cdot(q-q_{t}) +
\frac{1}{2} (q-q_{t})^{T} \cdot V^{\prime\prime} (q_{t}) \cdot(q-q_{t})
\label{eq:lha}%
\end{equation}
of the true potential $V(q)$ around the center of the wavepacket. In
Eq.~(\ref{eq:gwp}),
$q_{t}$ and $p_{t}$ are the expectation values of position
and momentum, $S_{t}$ is the classical action, and $Q_{t}$ and
$P_{t}$ are $D \times D$ complex matrices satisfying the
relations\cite{Hagedorn:1980,Hagedorn:1998,Faou_Lubich:2009,Begusic_Vanicek:2019,note:QP}
\begin{align}
Q_{t}^{T} \cdot P_{t} - P_{t}^{T} \cdot Q_{t}  &  = 0, \label{eq:QP_relation1}%
\\
Q_{t}^{\dagger} \cdot P_{t} - P_{t}^{\dagger} \cdot Q_{t}  &  = 2 i I,
\label{eq:QP_relation2}%
\end{align}
where $I$ is the $D \times D$ identity matrix. Without any further
approximation than the local harmonic approximation in Eq.~(\ref{eq:lha}), the
solution of the time-dependent Schr\"{o}dinger equation is equivalent to
propagating the Gaussian's parameters as\cite{Heller:1975, book_Lubich:2008,
Faou_Lubich:2009}
\begin{align}
\dot{q}_{t}  &  = m^{-1} \cdot p_{t}, \qquad \dot{p}_{t}   = - V^{\prime}(q_{t}),\label{eq:qp_t_dot}\\
\dot{Q}_{t}  &  = m^{-1} \cdot P_{t}, \qquad \dot{P}_{t}   = - V^{\prime\prime}(q_{t}) \cdot Q_{t}. \label{eq:QP_t_dot}
\end{align}

For Herzberg--Teller spectra,\cite{Herzberg_Teller:1933} where the transition
dipole moment is a linear function of position, the initial wavepacket,
\begin{equation}
\phi(q, 0) = [\mu(q_{0}) + \mu^{\prime} (q_{0}) ^{T} \cdot(q-q_{0})] \psi(q,
0), \label{eq:phi_0}%
\end{equation}
is no longer a simple Gaussian. Nevertheless, such a wavepacket also preserves
its form in the local harmonic potential (\ref{eq:lha}),
\cite{Lee_Heller:1982, Patoz_Vanicek:2018, Begusic_Vanicek:2018} namely
\begin{equation}
\phi(q, t) = [\mu(q_{0}) + \mu^{\prime} (q_{0}) ^{T} \cdot Q_{0} \cdot
Q_{t}^{-1} \cdot(q-q_{t})] \psi(q, t), \label{eq:phi_t}%
\end{equation}
where $\psi(q,t)$ is the Gaussian wavepacket (\ref{eq:gwp}) propagated with the
standard thawed Gaussian equations of motion for the parameters
[Eqs.~(\ref{eq:qp_t_dot})--(\ref{eq:QP_t_dot})]. This \emph{extended} thawed
Gaussian approximation has been recently applied to compute spectra beyond the
Condon approximation.\cite{Patoz_Vanicek:2018, Begusic_Vanicek:2018} In general, the Herzberg--Teller effect becomes important in weak or forbidden transitions, where the constant, Condon term of the transition dipole moment is small. However, it is hard to predict \textit{a priori} whether this effect contributes to the spectrum.

The thawed Gaussian approximation requires not only potential energies and
gradients but also Hessians at each point along the trajectory. This can become
rather costly for accurate \textit{ab initio} calculations of large
molecules.\cite{Wehrle_Vanicek:2014, Tatchen_Pollak:2009, Ceotto_Hase:2013, Zhuang_Ceotto:2013,
Ianconescu_Pollak:2013, Richings_Lasorne:2015, Richings_Worth:2015,
Alborzpour_Habershon:2016, Laude_Richardson:2018, Bonfanti_Pollak:2018, Polyak_Knowles:2019,
Conte_Ceotto:2019, Gabas_Ceotto:2019, Micciarelli_Ceotto:2019} For this
reason, two of us have proposed the \emph{single-Hessian} thawed Gaussian
approximation,\cite{Begusic_Vanicek:2019} where $V^{\prime\prime}(q_{t})$ of Eq.~(\ref{eq:QP_t_dot}) is replaced with
the reference Hessian $V^{\prime\prime}_{\text{ref}}(q_{\text{ref}})$ evaluated at a single (reference) point $q_{\text{ref}}$. The method was shown to perform well and consistently better than the standard global harmonic approaches in systems exhibiting moderate anharmonicity effects.\cite{Begusic_Vanicek:2019} Moreover, it provides an estimate of the effect of anharmonicity on spectra at little additional computational cost: compared to the global harmonic method, it requires in addition only a single \textit{ab initio} classical trajectory.

In Ref.~\citenum{Begusic_Vanicek:2019}, the single-Hessian thawed
Gaussian approximation was used only for Gaussian wavepackets (\ref{eq:gwp}). Here,
we combine the single-Hessian idea with the extended thawed Gaussian
approximation in order to accelerate calculations of Herzberg--Teller spectra.
Remarkably, Eq.~(\ref{eq:phi_t}) is unaffected with this change.
In contrast, the conservation of energy, derived for the single-Hessian thawed Gaussian
wavepacket in Ref.~\citenum{Begusic_Vanicek:2019}, does not hold in general for the extended thawed Gaussian wavepacket, for which the time derivative of the total energy is
\begin{equation}
\frac{dE}{dt}  = \hbar \text{Re} [ \mu(q_{0}) \mu^{\prime}(q_{0})^{T} \cdot Q_{0} \cdot
Q_{t}^{\dagger} \cdot b_{t}], \label{eq:de_dt}
\end{equation}
with $b_{t} :=(V^{\prime\prime}(q_{t}) - V_{\text{ref}}^{\prime\prime} (q_{\text{ref}} ) )\cdot m^{-1} \cdot p_{t}$ (see Supporting Information). Although the time derivative of energy (\ref{eq:de_dt}) is non-zero in general, the energy is conserved in purely Herzberg--Teller spectra, i.e., if the constant, Condon, term $\mu(q_{0})$ is zero.

\section{\label{sec:compdet}Computational and experimental details}

To estimate adiabaticity with the MSDR, the underlying nuclear
dynamics was based on Born-Oppenheimer dynamics, standard Tully's fewest-switches
surface hopping\cite{Tully:1990}, surface hopping with the energy-based decoherence correction
,\cite{Granucci_Persico:2007} or Ehrenfest dynamics. \textit{Ab initio}
trajectories were propagated using forces and NAC vectors obtained with CASSCF
electronic structure. However, to simulate vibrationally resolved spectra, it
is crucial to include dynamical correlation effects which are missing in
CASSCF. To avoid cumbersome CASPT2 \textit{ab initio} treatment, we employed
the second-order algebraic diagrammatic construction [ADC(2)] method, which includes important correlation effects for a balanced
treatment of so-called L$_{a}$ and L$_{b}$ states (S$_{2}$ and S$_{1}$ states
in azulene, respectively).\cite{Prlj_Corminboeuf:2016} A so-called ``adiabatic
Hessian'',\cite{AvilaFerrer_Santoro:2012, Begusic_Vanicek:2019} which is
evaluated at the optimized geometry of the final electronic state, was used as
the reference Hessian for the single-Hessian thawed Gaussian propagation. Dynamics and spectra simulations were performed with an in-house code coupled to Gaussian16,\cite{g16} Molpro2012,\cite{Werner_Schutz:2012,MOLPRO:2012} and Molpro2015\cite{MOLPRO2015,Kats_Schutz:2009} electronic structure packages. For
further details about dynamics simulations, electronic structure, and spectra
computations, see Supporting Information.

The absorption spectra were recorded using a PerkinElmer Lambda 950 UV/Vis/NIR
spectrophotometer in cyclohexane at room temperature with azulene concentration
of $10^{-5}$~M for S$_{2}$ spectrum and $10^{-3}$~M for the weaker S$_{1}$ band.
As for the emission, the spectra were recorded using a Horiba Jobin-Yvon 
Fluorolog-3 with a photomultiplier tube as a detector, the concentration was 
$10^{-5}$~M in cyclohexane, and the sample was excited at 280 nm.

\section{\label{sec:res}Results and discussion}

\subsection{\label{subsec:adiabaticity}Population dynamics and adiabaticity}

\begin{figure}[th]
\includegraphics{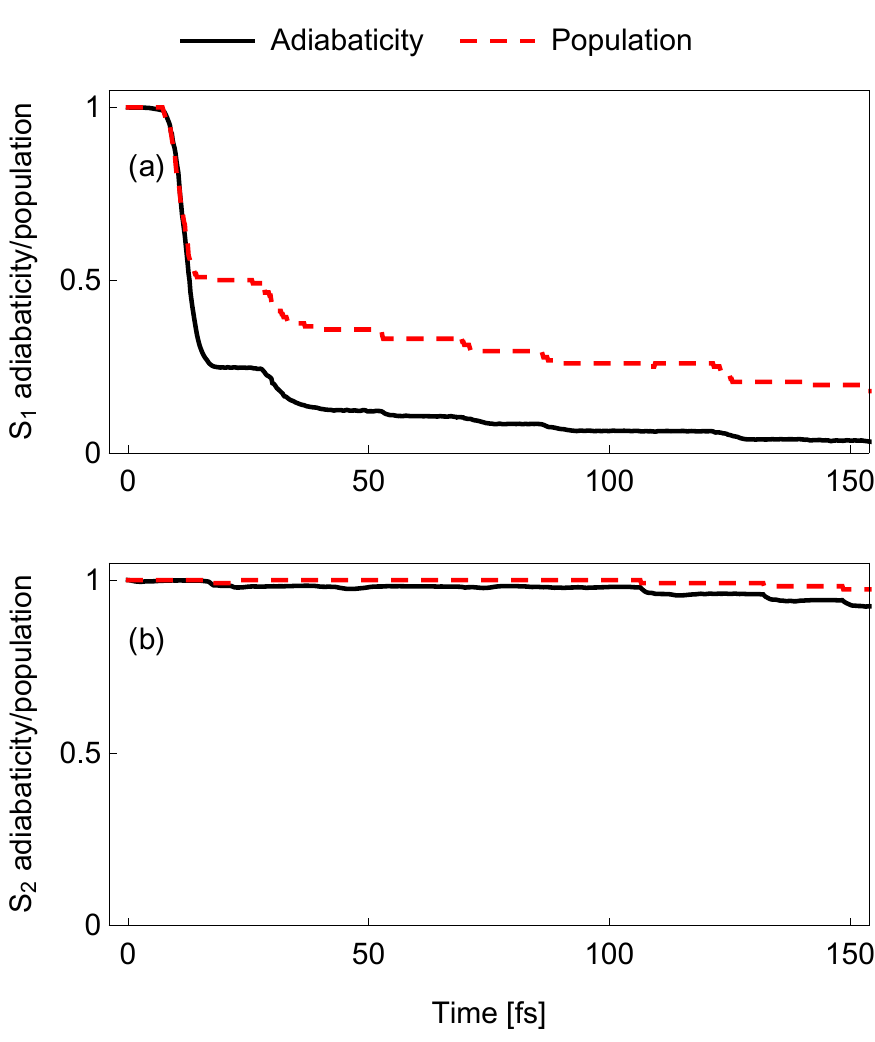}
\caption{Adiabaticity [Eqs.~(\ref{eq:adiab})--(\ref{eq:MSDR_c})] and population decay for an ensemble of trajectories initiated at: a) S$_1$ state or b) S$_2$ state, and evolved with the fewest switches surface hopping algorithm\cite{Tully:1990} with decoherence correction.\cite{Granucci_Persico:2007}}
\label{fig:Figure1}
\end{figure}

Nonadiabatic dynamics, approximated with the decoherence-corrected surface
hopping, was initiated in either the first or second excited state (Figure
\ref{fig:Figure1}). Subsequent populations of S$_{1}$ and S$_{2}$ states
illustrate well the violation of Kasha's rule in azulene. On one hand, the
system excited to S$_{1}$ decays quickly to the ground state due to the
accessible conical intersection seam. On the other hand, the system excited to
S$_{2}$ remains in that state, indicating that nonradiative decay is negligible.
Interestingly, the S$_{1}$ population decay appears as at least a biexponential process, where only the slower time constant is comparable to experiments.\cite{Diau_Zewail:1999}

Despite the appealing picture provided by the population analysis, populations
alone are not sufficient to account for all non-Born-Oppenheimer effects,
including the subtle effects of wavepacket displacement and interferences, including geometric phase,
induced by NACs, even on a single potential energy
surface.\cite{Zimmermann_Vanicek:2012, Xie_Yarkony:2019} Adiabaticity is, indeed, a more
rigorous way to evaluate the importance of NACs. The MSDR, in turn, makes it
possible to estimate adiabaticity with little additional computational cost.
As shown in Fig.~\ref{fig:Figure1}, S$_{1}$ adiabaticity significantly drops
already after 10~fs, which corresponds to the first arrival of the wavepacket
to the conical intersection region, and gradually approaches zero within
150~fs. In the same time interval, S$_{2}$ adiabaticity remains quite high.
Overall, the computed adiabaticity
provides additional support for the disparate behaviors of S$_{1}$ and S$_{2}%
$. Interestingly, the S$_1$ adiabaticity computed with the simple Born-Oppenheimer
dynamics (see Fig.~S2), which contains no information about populations whatsoever,
resembles that of Fig.~\ref{fig:Figure1}a. In contrast, mean-field Ehrenfest dynamics and standard surface hopping (without decoherence correction) yield higher adiabaticity of dynamics started from the S$_1$ state and lower adiabaticity of dynamics started from S$_2$; similar trends are observed for the initial-state populations (see Figs.~S2--S5).

\subsection{\label{subsec:spec_azulene}Absorption and emission spectra of azulene}

\begin{figure}[pth]
\includegraphics{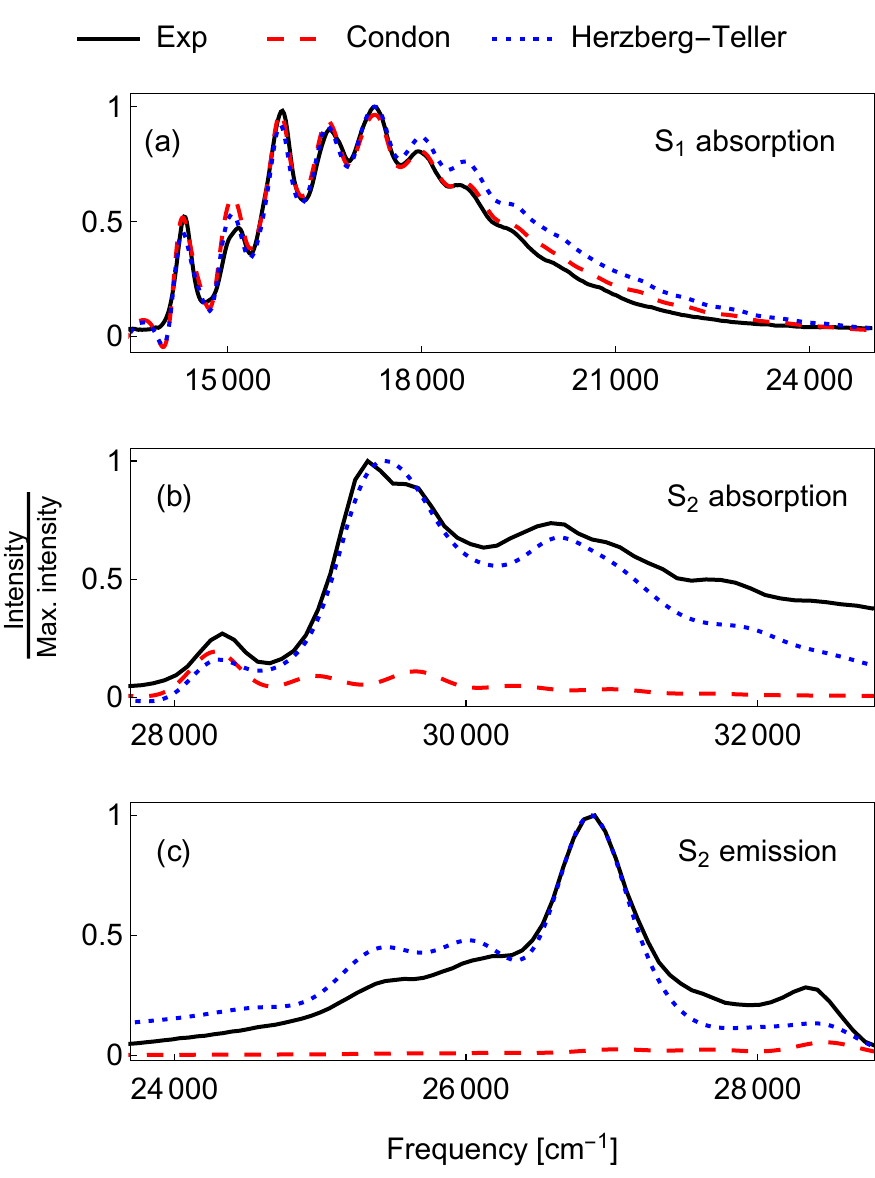} \caption{\label{fig:OTF_Paper}Vibrationally resolved (a) $\text{S}_1 \leftarrow \text{S}_0$ absorption, (b) $\text{S}_2 \leftarrow \text{S}_0$ absorption, and (c)  $\text{S}_2 \rightarrow \text{S}_0$ emission spectra of azulene. Calculations using adiabatic single-Hessian thawed Gaussian approximation (see Sec.~\ref{subsec:tga} and Table~S1) for the wavepacket dynamics and either Condon [$\mu(q) \approx \mu(q_0)$] or Herzberg--Teller [Eq.~(\ref{eq:phi_0})] approximations for the transition dipole moment are compared with the experiment. To facilitate this comparison, all computed spectra are shifted in frequency by a constant (see Table~S2) and are rescaled to unit maximum intensity, except for those computed within the Condon approximation, which are scaled by the maxima of the corresponding Herzberg--Teller spectra.}
\end{figure}

Both population dynamics and adiabaticity suggest that the dynamics of a
wavepacket initially in the S$_{2}$ electronic state can be described rather
well within the Born--Oppenheimer approximation, unlike the dynamics started
in the S$_{1}$ state, which exhibits fast nonradiative decay to the S$_{0}$
ground state. Therefore, one would expect the thawed Gaussian approximation, a
method that neglects nonadiabatic effects, to perform better for the
$\text{S}_{2} \leftarrow\text{S}_{0}$ absorption spectrum than for the
$\text{S}_{1} \leftarrow\text{S}_{0}$ absorption spectrum.

\begin{figure}[pth]
\includegraphics{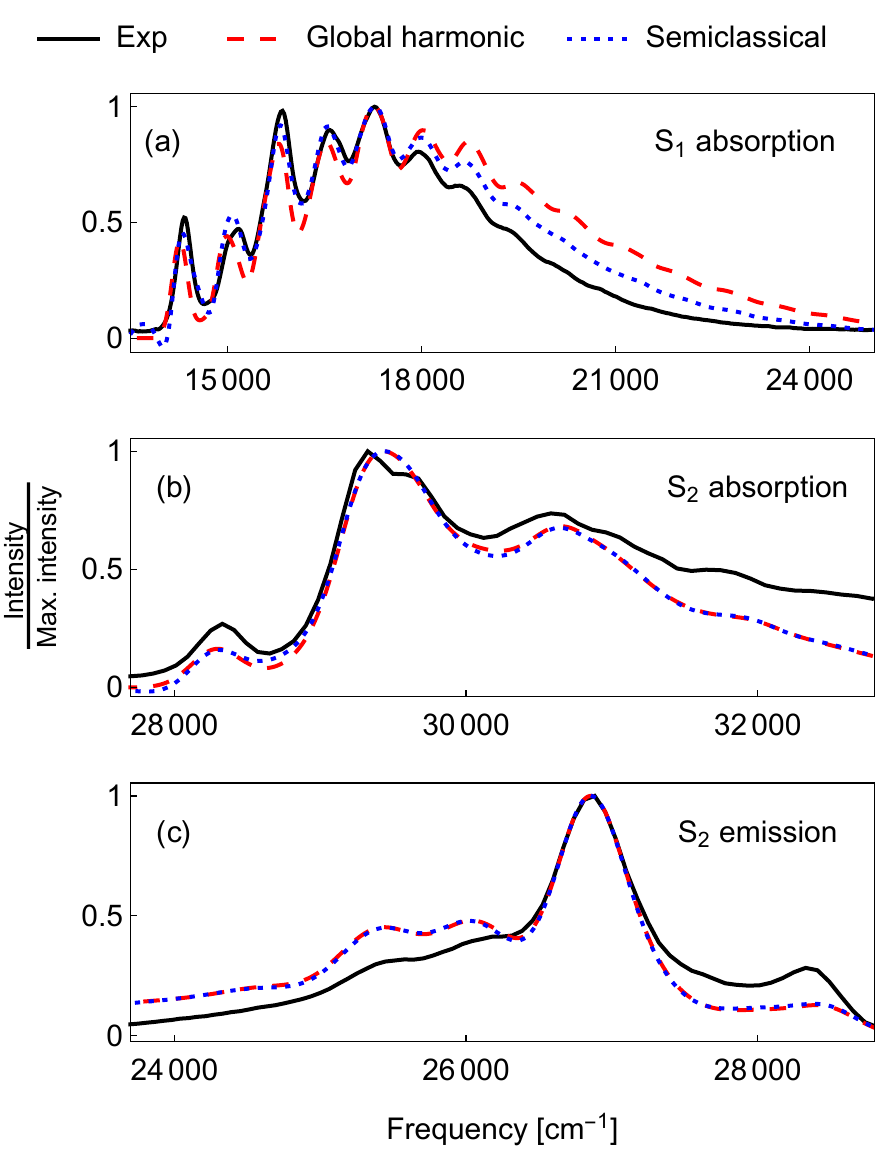}
\caption{\label{fig:OTFvsGH_Paper}Vibrationally resolved (a) $\text{S}_1 \leftarrow \text{S}_0$ absorption, (b)
$\text{S}_2 \leftarrow \text{S}_0$ absorption, and (c)
$\text{S}_2 \rightarrow \text{S}_0$ emission spectra of azulene. Calculations using adiabatic single-Hessian thawed Gaussian approximation (``semiclassical,'' see Sec.~\ref{subsec:tga} and Table~S1) or adiabatic global harmonic approach (as described in Ref.~\citenum{Santoro_Jacquemin:2016})---both combined with the Herzberg--Teller approximation [Eq.~(\ref{eq:phi_0})] for the transition dipole moment---are compared with the experiment. To facilitate this comparison, all computed spectra are rescaled to unit maximum intensity and shifted in frequency by a constant (see Table~S2).}
\end{figure}

Surprisingly, the simulated S$_{1} \leftarrow\text{S}_{0}$ absorption spectrum
(see Fig.~\ref{fig:OTF_Paper}a) agrees rather well with the experiment. It
appears that, despite being considerably different from the true
nonadiabatically evolved wavepacket, the thawed Gaussian wavepacket results in
a correct autocorrelation function. Since only the part of the wavepacket
that remains on the initial state contributes to the autocorrelation function
(\ref{eq:C_t}), a more convenient measure of nonadiabatic effects on spectra is obtained
by dividing the adiabaticity by population. Adiabaticity is equal to the initial-state 
population when the nonadiabatic coupling affects only the amplitude but not the shape 
of the nuclear wavepacket on the initial surface. The ratio between the adiabaticity and initial-state
population, shown in Fig.~S6, decays less dramatically than the adiabaticity, which
justifies partially the accuracy of the spectra computed using
the Born--Oppenheimer wavepacket dynamics. In addition, rather short times are needed for the computation of the spectrum because it is only moderately resolved. One could expect the wavepacket autocorrelation function to exhibit increasingly more nonadiabatic effects at later times, implying that these effects would have to be included in the simulation of the high-resolution absorption spectrum. As already reported in
Refs.~\citenum{Dierksen_Grimme:2004} and \citenum{Niu_Shuai:2010}, the S$_{1}$
absorption spectrum can be computed easily within the Condon approximation and
even using global harmonic models. Nevertheless, we observe an
improvement of the computed spectrum by using the on-the-fly thawed Gaussian
method that partially accounts for anharmonicity (see
Fig.~\ref{fig:OTFvsGH_Paper}a); including the Herzberg--Teller contribution, however,
does not improve the spectrum (Fig.~\ref{fig:OTF_Paper}a).

S$_{2}$ absorption and emission spectra are also well described by the
single-Hessian extended thawed Gaussian approximation. The corresponding
potential energy surface is harmonic in the regions visited by the nuclear
wavepacket, which is confirmed by comparing spectra computed with thawed
Gaussian and global harmonic approaches (see Figs.~\ref{fig:OTFvsGH_Paper}b
and \ref{fig:OTFvsGH_Paper}c). In contrast to the S$_1$ spectrum, for 
describing the S$_2$ spectra, the Herzberg--Teller contribution due to 
coupling with higher excited electronic states\cite{Li_Lin:2010,
Patoz_Vanicek:2018, Begusic_Vanicek:2018} is essential (see Figs.~\ref{fig:OTF_Paper}b 
and \ref{fig:OTF_Paper}c). This effect has only been analyzed qualitatively in the
emission spectrum of azulene, but never in the S$_{2}$ absorption
spectrum.\cite{Gustav_Storch:1990}

\begin{figure}[pth]
\includegraphics{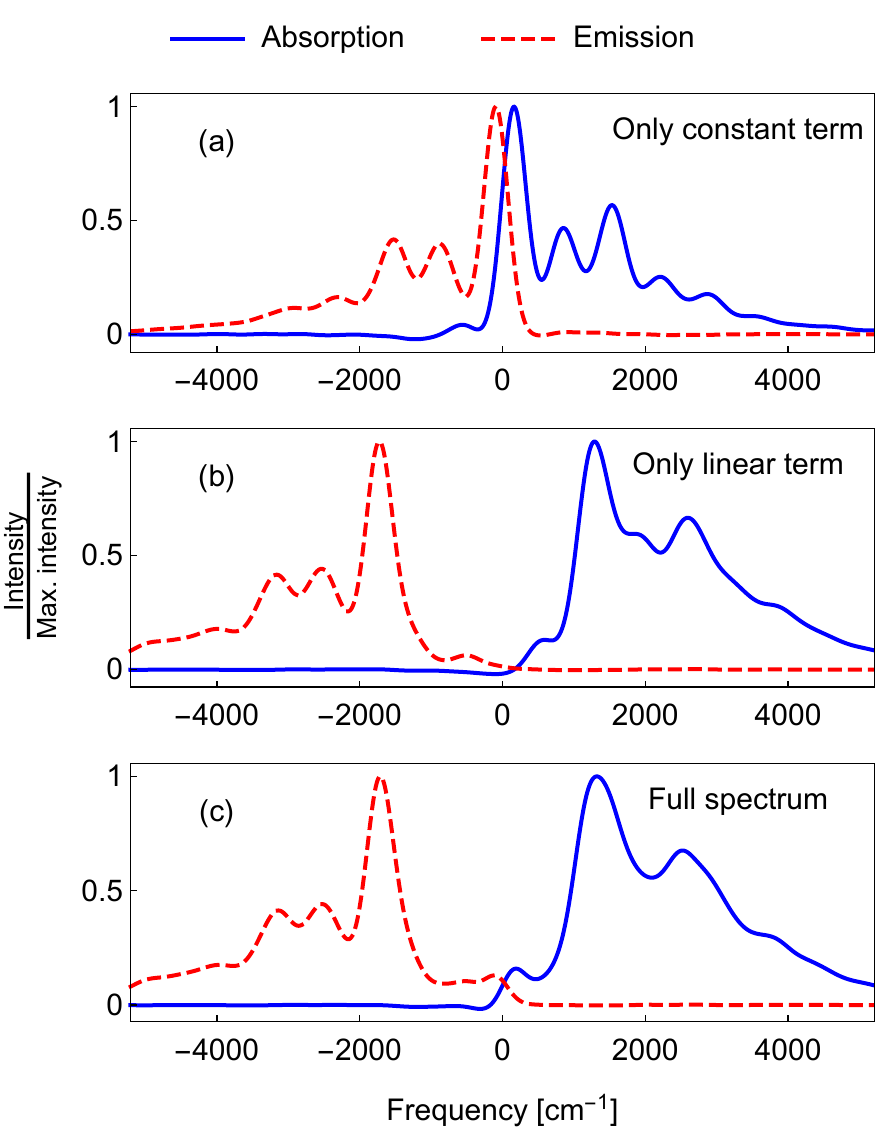} \caption{\label{fig:MirrorImageS2}Vibrationally resolved S$_2$ absorption and emission spectra of azulene computed with (a) only Condon (constant) term, (b) only Herzberg--Teller (linear) term, (c) both Condon and Herzberg--Teller terms in the expansion of the transition dipole moment  [with all calculations using the adiabatic single-Hessian (extended) thawed Gaussian approximation (see Sec.~\ref{subsec:tga} and Table~S1)]. To facilitate comparison between absorption and emission, all spectra are rescaled to unit maximum intensity and shifted in frequency by a constant.}
\end{figure}

Furthermore, the Herzberg--Teller coupling is responsible for the breakdown of
mirror image symmetry between the absorption and emission spectra, which is
formally valid only for a displaced harmonic oscillator model within the
Condon approximation.\cite{Santoro_Barone:2008,Pour_Hauer:2017} In general,
changes in the force constant, mode coupling, anharmonicity, and
Herzberg--Teller coupling can all break this symmetry. In
Fig.~\ref{fig:MirrorImageS2}a, we show that the Condon absorption and emission
spectra retain (to a large extent) this symmetry, whereas the mirror image
symmetry is broken completely in the case of Herzberg--Teller
spectra (see Fig.~\ref{fig:MirrorImageS2}c). Such effect of the
Herzberg--Teller coupling is well
known,\cite{Geigle_Hohlneicher:1997,Santoro_Barone:2008, Toutounji:2019,
Toutounji:2019a} but is commonly interpreted in terms of the cross terms that
arise when both Condon and Herzberg--Teller contributions to the spectrum are
significant. This is not the case here, as significant asymmetry is observed
even for the pure Herzberg--Teller contribution (where the constant Condon
term is set to zero, see Fig.~\ref{fig:MirrorImageS2}b). In azulene, the breakdown of the mirror image symmetry
between absorption and emission is a result of an interplay between the
Herzberg--Teller and mode-mixing (Duschinsky) effects. Indeed, the symmetry is
mostly recovered if either of the two effects is ``turned off''
(see Fig.~\ref{fig:MirrorImageS2}a, where Herzberg--Teller coupling is set to
zero, and Fig.~S7, where mode mixing is neglected). More precisely, coupling between the modes modifies only slightly the dynamics of the Gaussian wavepacket, hence the similarity between the spectra in Figs.~\ref{fig:MirrorImageS2}a and S7a, but affects considerably the linear, Herzberg--Teller term of the extended thawed Gaussian wavepacket (\ref{eq:phi_t}), which explains the difference between the spectra in Figs.~\ref{fig:MirrorImageS2}b and S7b. The Duschinsky effect on the absorption spectrum is largely due to couplings between the Herzberg--Teller active modes (see Fig.~S8 where only those couplings are neglected). In contrast, the emission spectrum is only weakly affected by the mode-mode couplings (compare the emission spectra in Figs.~\ref{fig:MirrorImageS2}c and S7c).

Small discrepancies between the simulated spectra and experiments are likely
due to the accuracy of the electronic structure method. We found that
the accuracy of the computed S$_{2} \leftarrow$ S$_0$ absorption spectrum depends strongly on
the degree of dynamic correlation included in the \textit{ab initio} method
(see Fig.~S1). Accounting for finite-temperature and solvent effects, which
are in our calculations included only phenomenologically through Gaussian
broadening, might further improve the accuracy.\cite{Baiardi_Barone:2013,
Reddy_Prasad:2016, Borrelli_Gelin:2016, Chen_Zhao:2017, Chen_Tanimura:2015, Loco_Mennucci:2018, Loco_Cupellini:2019,
Fortino_Pedone:2019}

\section{\label{sec:conclusion}Conclusion}

To conclude, we presented a systematic and general semiclassical approach for studying photophysics beyond Kasha's rule and spectroscopy beyond Condon's approximation. We validated the method on the challenging case of azulene, where the proposed approach allowed us to consider the interplay of nonadiabatic, anharmonicity, mode-mixing, and Herzberg--Teller effects, as well as the importance of
dynamical electron correlation in the electronic structure methods used. The
presented methodology allows one to perform in-depth studies of photochemistry
and photophysics of various molecular systems at a moderate computational cost.

\begin{acknowledgement}
The authors acknowledge the financial support from the Swiss National Science Foundation through the NCCR 
MUST (Molecular Ultrafast Science and Technology) Network and from the European Research Council (ERC) under the European Union's Horizon 2020 research and innovation programme (grant agreement No. 683069 -- MOLEQULE).
\end{acknowledgement}

\begin{suppinfo}
Computational details, comparison of electronic structure methods for spectra calculations, energy conservation for the single-Hessian extended thawed Gaussian wavepacket, adiabaticity and population dynamics for different nuclear dynamics methods, adiabaticity divided by the initial-state population, S$_2$ absorption and emission spectra simulated without Duschinsky mixing of the normal modes.
\end{suppinfo}

\bibliography{biblio45, additions_Azulene}

\end{document}


\graphicspath{{"d:/Group Vanicek/Desktop/Azulene/figures/"}
{./figures/}{C:/Users/Jiri/Dropbox/Papers/Chemistry_papers/2018/SingleHessian/figures/}}

\newpage
\section{\label{sec:compdet_SI}Computational details}
Minima of S$_0$, S$_1$, and S$_2$ states, as well as minimum energy conical intersection between S$_0$ and S$_1$ were optimized at SA5-CASSCF(4,6)/6-31G* level. To compute the potential energy surface cuts in Fig.~1, we performed linear interpolation of internal coordinates between the optimized geometries, while the section of the surface beyond the S$_2$ minimum was based on internal coordinate extrapolation.
All dynamics simulations were performed for 800 time steps of 8 a.u. (0.1935 fs) each, therefore, for the total time of $\approx 155$~fs. The velocity Verlet algorithm was used to integrate classical equations of motion.

\subsection{\label{subsec:compdet_SI_MSDR}Multiple-surface dephasing representation (MSDR)}
Ensembles of $N=112$ \textit{ab initio} trajectories were propagated with each nuclear dynamics method (i.e., Born-Oppenheimer, fewest-switches surface hopping, or Ehrenfest dynamics). Note that by ``Ehrenfest dynamics'' we mean a locally mean field dynamics,\cite{Zimmermann_Vanicek:2012a} i.e., an independent Ehrenfest dynamics of each trajectory in the ensemble. For nonadiabatic dynamics, ground and four excited states were taken into account. Energies, gradients, and nonadiabatic couplings were computed with state-averaged complete active space self-consistent field [SA5-CASSCF(4,6)/6-31G*] electronic structure method, as implemented in Molpro2012.\cite{Werner_Schutz:2012,MOLPRO:2012} Compared to CASSCF with larger active spaces [(6,6),(10,10)], which give an incorrect ground state minimum structure with C$_s$ symmetry, CASSCF(4,6) has a correct C$_{2v}$ minimum. Initial positions and momenta were sampled from the Wigner distribution of a vibrational ground state of a harmonic fit to the ground potential energy surface. Assuming the vertical excitation, the whole ensemble of trajectories was launched from either the S$_1$ or S$_2$ state. Surface hopping simulations were performed both without\cite{Tully:1990} and with an energy-based decoherence correction,\cite{Granucci_Persico:2007} which was applied at every nuclear time step, with a parameter $\alpha=0.1$ Hartree.

\subsection{\label{subsec:compdet_SI_TGA}On-the-fly \textit{ab initio} thawed Gaussian propagation and spectra}
\begin{figure}[ht]
\includegraphics{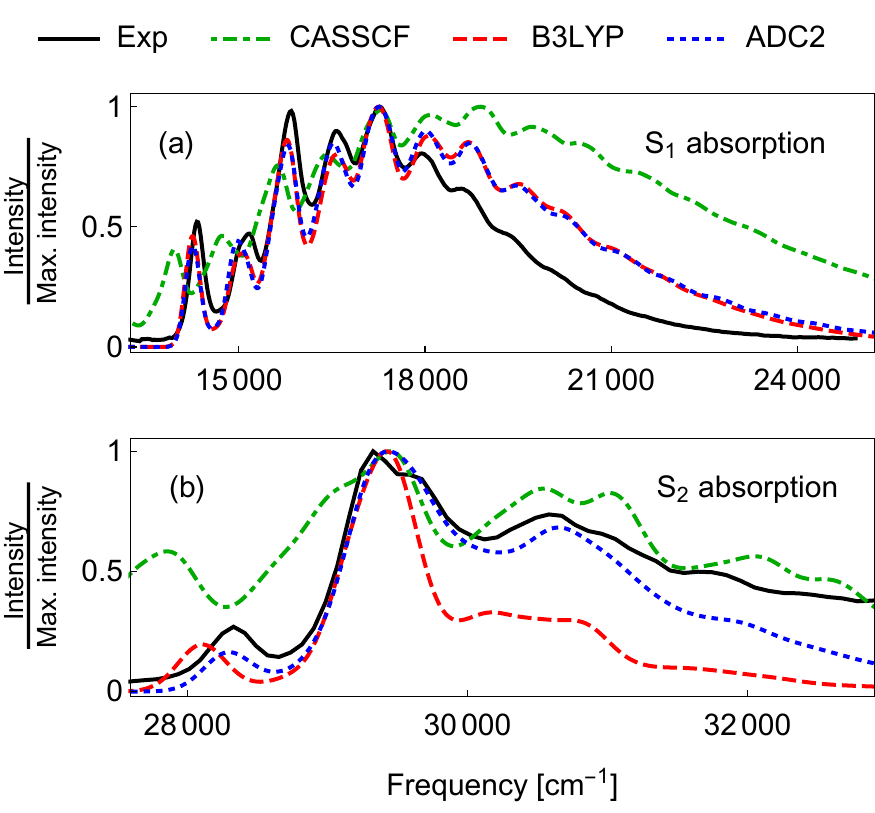}
\caption{{\label{fig:ElectronicStructureComparison_Paper}}S$_1 \leftarrow$ S$_0$ and S$_2\leftarrow$ S$_0$ absorption spectra computed using different electronic structure methods.} 
\end{figure}

For computing vibronic spectra, three electronic structure methods were tested:
(i) (time-dependent) density functional theory (in combination with the B3LYP functional and TZVP basis set, which were used in Ref.~\citenum{Dierksen_Grimme:2004}); 
(ii) SA5-CASSCF(4,6)/6-31G* (used also for MSDR calculations);
(iii) second-order M{\o}ller--Plesset perturbation theory (MP2) for ground state combined with the second-order algebraic diagrammatic construction [ADC(2)] scheme for the excited states (cc-pVDZ basis set). Gaussian 16 package\cite{g16} was used for (time-dependent) density functional theory calculations, while Molpro2015\cite{MOLPRO2015} was used for CASSCF, MP2, and ADC(2) methods. For ADC(2) calculations, we used Laplace transformed density-fitted local ADC(2) implementation in Molpro [keyword LT-DF-LADC(2)].\cite{Kats_Schutz:2009} Spectra computed with these electronic structure methods---including the Herzberg--Teller term of the transition dipole moment, but only within the adiabatic global harmonic approximation for the potential energy---are compared in Fig.~\ref{fig:ElectronicStructureComparison_Paper}.

The trajectories needed for evaluating the S$_1\leftarrow$ S$_0$ and S$_2\leftarrow$ S$_0$ absorption spectra with the single-Hessian extended thawed Gaussian approximation were propagated with the excited-state ADC(2) gradients, starting from the ground-state geometry optimized at the MP2 level. Similarly, the trajectory needed for computing the S$_2 \rightarrow$ S$_0$ emission spectrum started at the S$_2$ minimum [found by geometry optimization at the ADC(2) level of theory] and was run using MP2 gradients of the ground-state potential energy surface. In all cases, the initial wavepacket was the ground vibrational state of the harmonic potential fit to the potential energy surface of the initial electronic state. Gaussian wavepacket propagation was performed in normal-mode coordinates obtained by diagonalizing the mass-scaled initial-state Hessian, so that the initial wavepacket was a simple product of one-dimensional Gaussian functions. For each single-Hessian thawed Gaussian propagation, the reference Hessian was chosen as the adiabatic Hessian of the final electronic state (see Table~\ref{tab:comp_det}).

\begin{table}[ht]
    \caption{\label{tab:comp_det}Parameters used in the calculations of various spectra with the adiabatic single-Hessian extended thawed Gaussian approximation.}
    \begin{tabular}{lccc}
    \hline \hline
      &  S$_1$ absorption & S$_2$ absorption & S$_2$ emission\\
      Initial geometry $q_0$ &  $q_{\text{eq}}(\text{S}_0)$ & $q_{\text{eq}}(\text{S}_0)$ & $q_{\text{eq}}(\text{S}_2)$ \\
      Reference Hessian $V^{\prime \prime}_{\text{ref}}$ & $V^{\prime \prime}_{\text{S}_1}$ & $V^{\prime \prime}_{\text{S}_2}$ & $V^{\prime \prime}_{\text{S}_0}$ \\
      Reference geometry $q_{\text{ref}}$ &  $q_{\text{eq}}(\text{S}_1)$ & $q_{\text{eq}}(\text{S}_2)$ & $q_{\text{eq}}(\text{S}_0)$ \\
      \hline \hline
    \end{tabular}
\end{table}

Derivatives of the electronic transition dipole moment with respect to nuclear coordinates are not readily available in quantum chemistry packages. We evaluated them by finite differences, i.e., by computing the transition dipole moments at geometries displaced by 0.01 a.u. from the optimized initial-state geometry. Fortunately, these numbers can be extracted from the output of the \textit{ab initio} numerical excited-state force or Hessian calculation.\cite{Baiardi_Barone:2013,Patoz_Vanicek:2018, Begusic_Vanicek:2018}

Spectral broadening was introduced by multiplying the autocorrelation function with a Gaussian damping function, which is equivalent to convolving the spectrum with another, but related Gaussian function. To facilitate the comparison between computed and experimental spectra, we introduced a constant energy shift in each spectrum. Because the constant shift error arises mostly due to the incorrect \textit{ab initio} vertical energy gap, the same shifts were applied to Condon and Herzberg--Teller spectra. Broadening and energy shift parameters are given in Table~\ref{tab:hwhm_shift}.

\begin{table}[ht]
    \centering
    \caption{\label{tab:hwhm_shift}Half-width at half-maximum (HWHM) of the Gaussian broadening functions and horizontal energy shifts  applied to spectra computed with global harmonic models or thawed Gaussian approximation (TGA). All values are expressed in cm$^{-1}$. Exceptionally, in Fig.~5 of the main text and Fig.~\ref{fig:MirrorImageS2_withoutDuschinsky}, HWHM was 200~cm$^{-1}$ for both S$_2$ absorption and emission spectra.}
    \begin{tabular}{lccc}
        \hline \hline
      &  S$_1$ absorption & S$_2$ absorption & S$_2$ emission\\
      HWHM &  140 & 200 & 250 \\
      Energy shift (Global harmonic) & -5470 & -6230 & -6050 \\
      Energy shift (TGA) & -5740 & -6480 & -6040 \\
      \hline \hline
    \end{tabular}
\end{table}

\section{Energy non-conservation in the single-Hessian extended \\ thawed Gaussian approximation}

In the single-Hessian thawed Gaussian approximation,\cite{Begusic_Vanicek:2019} the potential energy is
approximated along the trajectory as
\begin{equation}
V_{\text{SH}} (q, t) = V(q_{t}) + V^{\prime}(q_{t}) \cdot(q - q_{t}) +
\frac{1}{2} (q - q_{t})^{T} \cdot V_{\text{ref}}^{\prime\prime} (q_{\text{ref}%
}) \cdot(q - q_{t}). \label{eq:shlha}%
\end{equation}
The time derivative of the total energy (based on $V_{\text{SH}}$) of the extended thawed Gaussian wavepacket is 
\begin{align}
\frac{dE}{dt}  &  = \frac{d}{dt} \langle\phi(t)| \hat{H}_{\text{eff}}(t)|\phi(t)\rangle\label{eq:de_dt_a}\\
&  =\langle\phi(t)|\frac{d}{dt}\hat{V}_{\text{SH}}
(t)|\phi(t)\rangle\label{eq:de_dt_b}\\
&  =\langle\phi(t)|b_{t}^{T}\cdot(\hat{q}-q_{t})|\phi(t)\rangle
\label{eq:de_dt_c}\\
&  =2 \text{Re} \{ \mu(q_{0}) \langle\psi(t)| \mu^{\prime}(q_{0})^{T} \cdot
Q_{0} \cdot Q_{t}^{-1} \cdot[(\hat{q}-q_{t} ) \otimes(\hat{q}-q_{t})^{T}]
\cdot b_{t} |\psi(t)\rangle\}\label{eq:de_dt_d}\\
&  = \hbar \text{Re} [ \mu(q_{0}) \mu^{\prime}(q_{0})^{T} \cdot Q_{0} \cdot
Q_{t}^{\dagger} \cdot b_{t}], \label{eq:de_dt_e}%
\end{align}
where $b_{t} :=
(V^{\prime\prime}(q_{t}) - V_{\text{ref}}^{\prime\prime} (q_{\text{ref}} ) ) \cdot m^{-1} \cdot p_{t}$.
Equation (\ref{eq:de_dt_b}) follows because
the thawed Gaussian solves exactly the Schr\"{o}dinger equation with
$\hat{H}_{\text{eff}} =  \frac{1}{2} \hat{p}^T \cdot m^{-1} \cdot \hat{p} + \hat{V}_{\text{SH}}(t)$. In Eq.~(\ref{eq:de_dt_d}) we used the fact that the
Gaussian probability density $|\psi(q,t)|^{2}$ is an even function centered at
$q_{t}$, i.e., the integrals of terms that are linear and cubic in $(q -
q_{t})$ vanish and in Eq.~(\ref{eq:de_dt_e}) we used the relation
$\langle\psi(t)|(\hat{q}-q_{t})\otimes(\hat{q}-q_{t})^{T}|\psi(t)\rangle=(\hbar / 2)
Q_{t} \cdot Q_{t}^{\dagger}$ for the position variance in state $\psi$.

\section{Adiabaticity and population dynamics evaluated with different nuclear dynamics methods}

\begin{figure}[ht]
\centering
\includegraphics{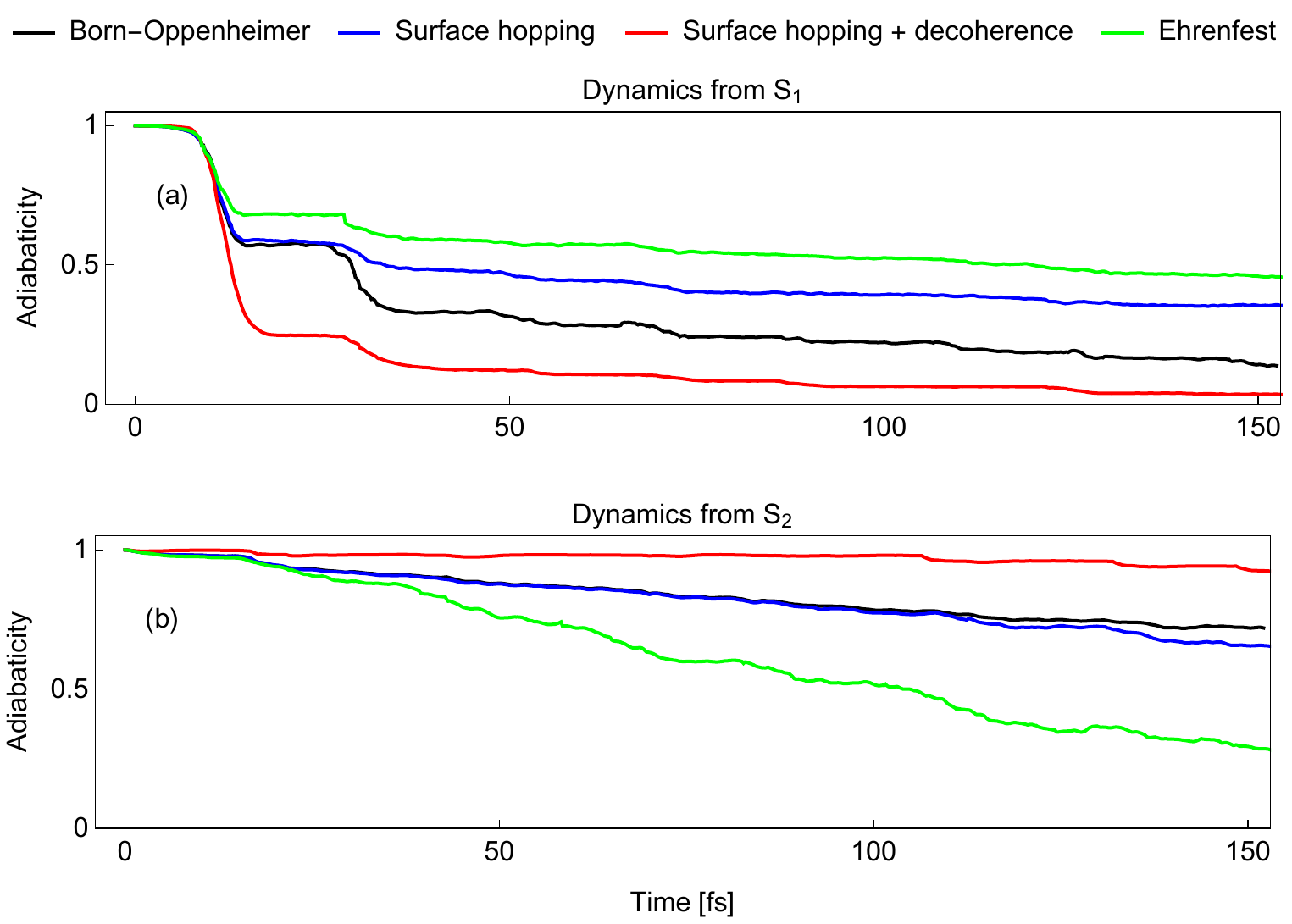}
\caption{Adiabaticity of the quantum dynamics initiated at the S$_1$ (upper panel) or S$_2$ (lower panel) state. Several nuclear dynamics methods are compared.}
\end{figure}

\begin{figure}[ht]
\centering
\includegraphics{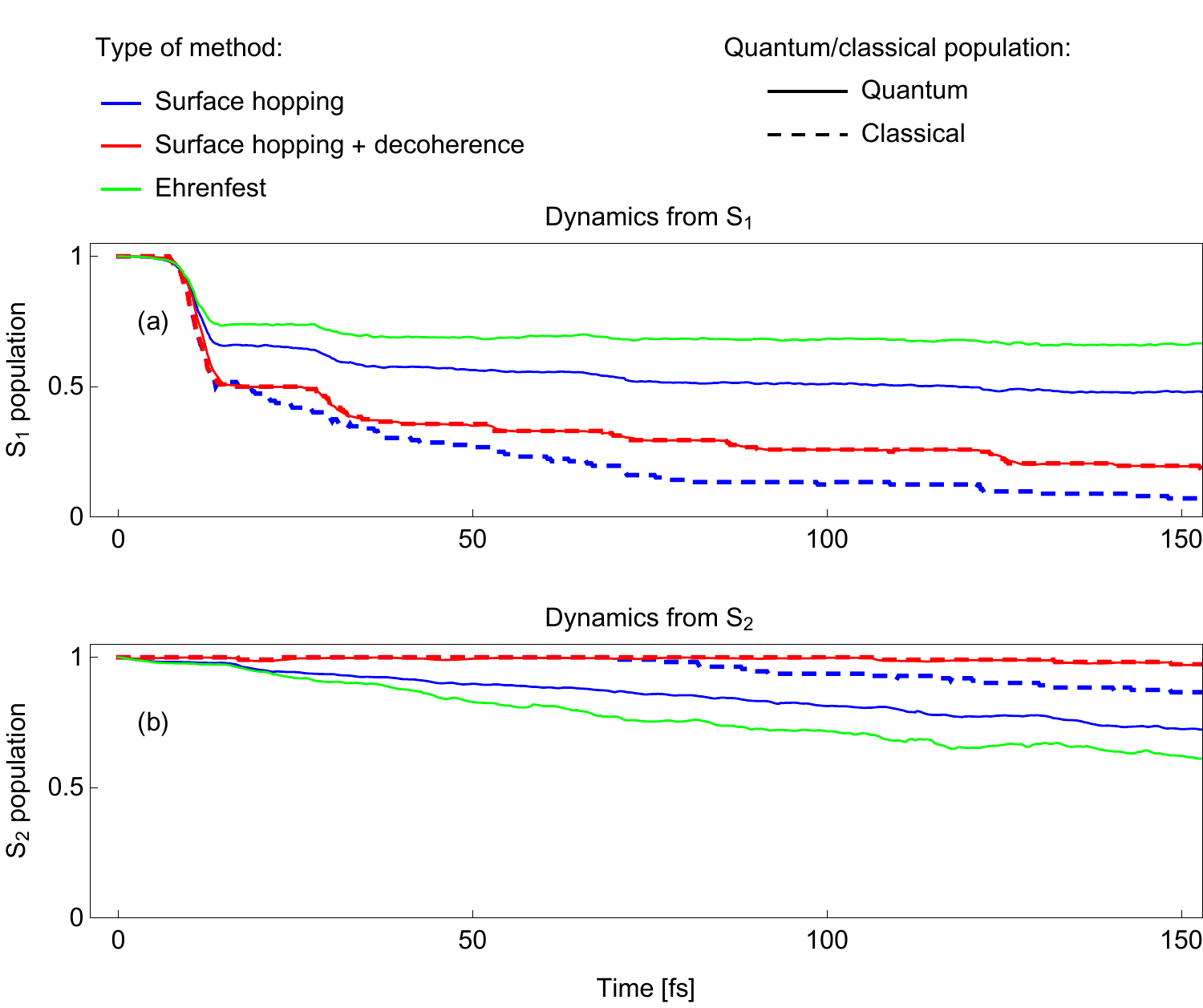}
\caption{Population of the S$_1$ state during the dynamics initiated at S$_1$ (upper panel) and population of the S$_2$ state during the dynamics initiated at S$_2$ (lower panel) computed with different nuclear dynamics methods.}
\end{figure}

\begin{figure}[ht]
\centering
\includegraphics[width=\textwidth]{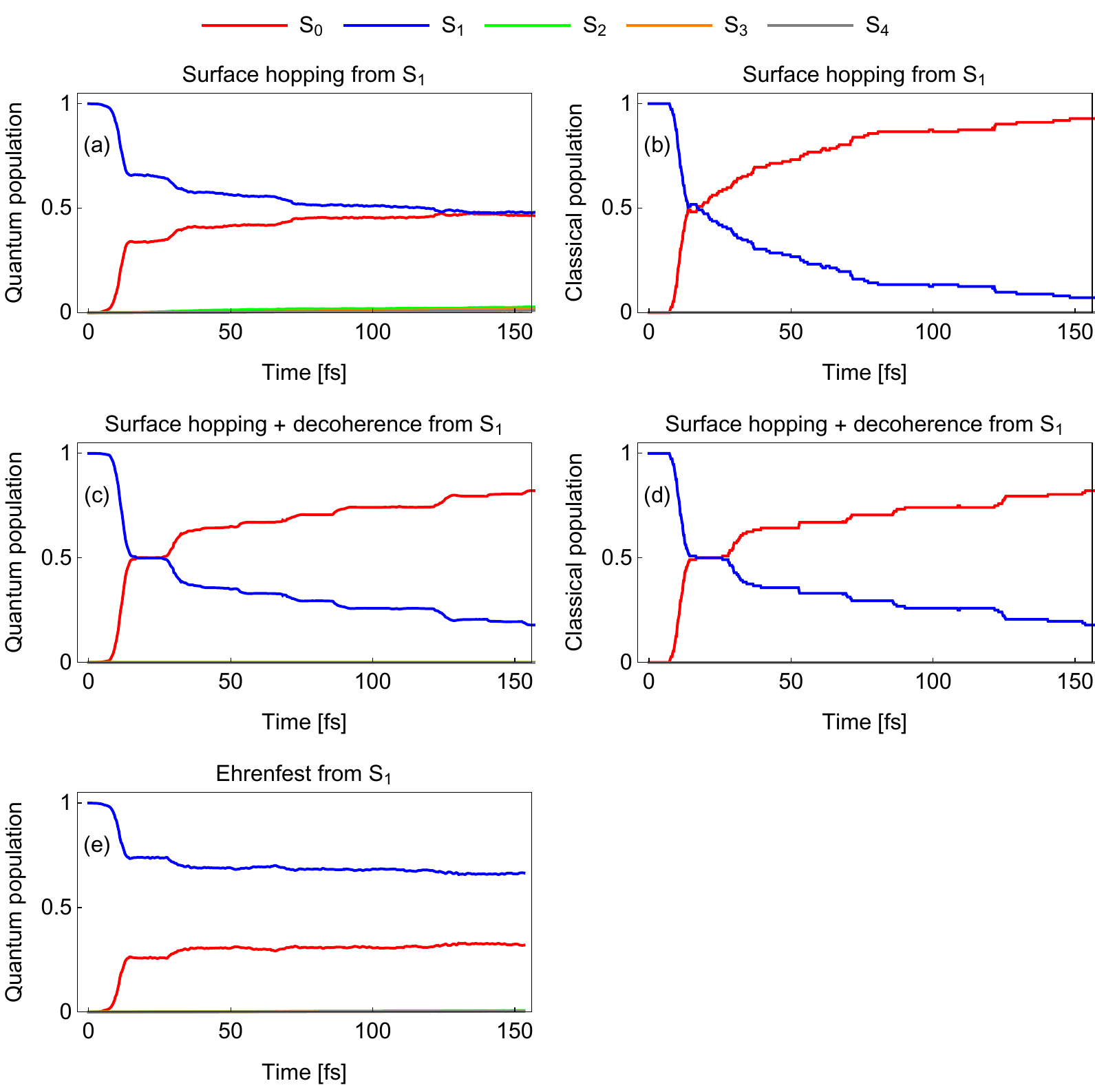}
\caption{Populations of states S$_0$--S$_4$ obtained with different nuclear dynamics simulations initiated at S$_1$.}
\end{figure}

\begin{figure}[ht]
\centering
\includegraphics[width=\textwidth]{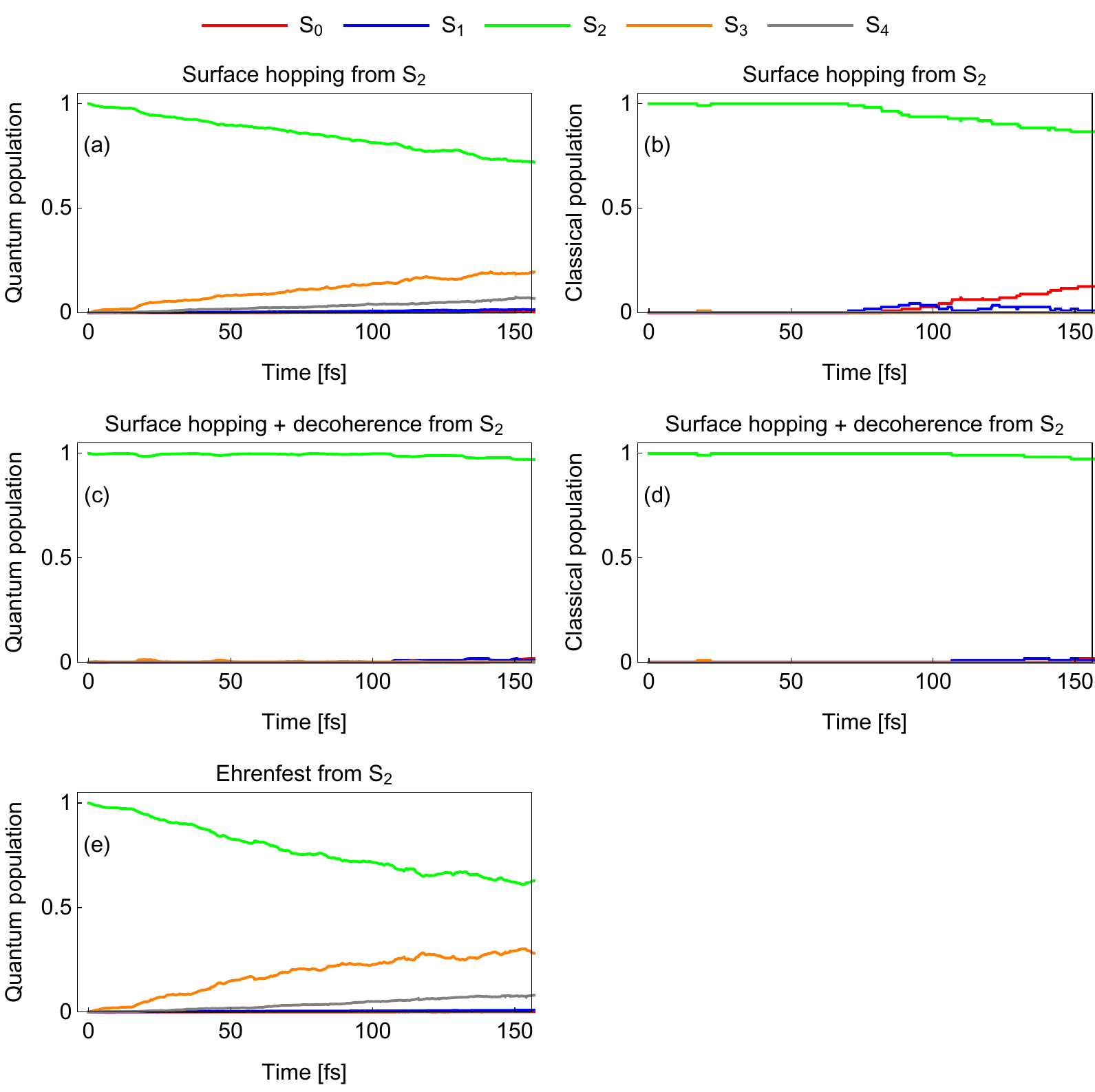}
\caption{Populations of states S$_0$--S$_4$ obtained with different nuclear dynamics simulations initiated at S$_2$.}
\end{figure}

\clearpage
\section{\label{sec:adiab_over_pop}Adiabaticity divided by initial-state population}
\begin{figure}[ht]
\includegraphics{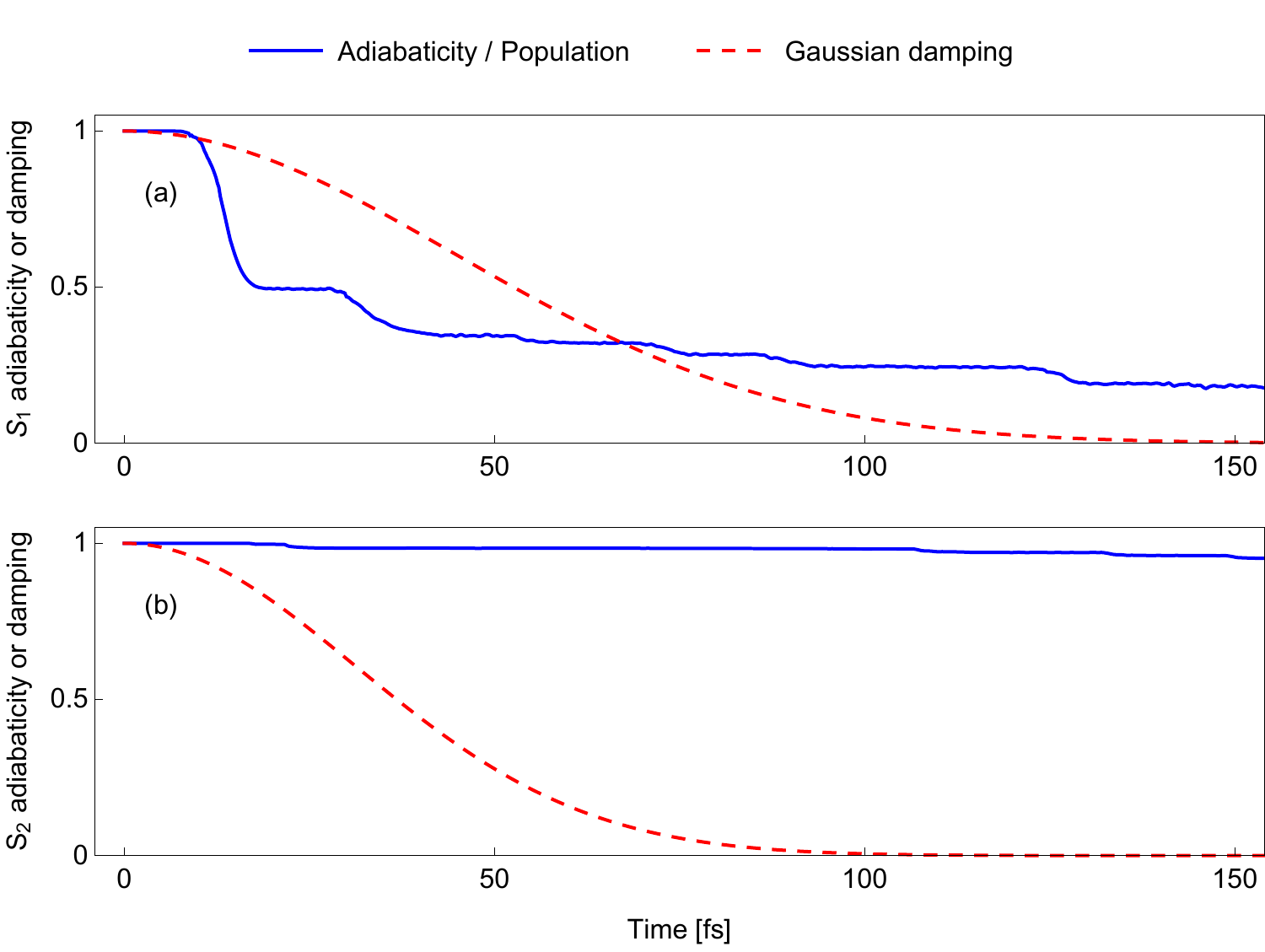}
\caption{\label{fig:FigureS5}Adiabaticities divided by the initial-state populations (both computed using surface hopping dynamics with decoherence correction, i.e. results shown in Fig.~2 of the main text) for the dynamics started at: (a) S$_1$, (b) S$_2$. Gaussian decay functions used for broadening the corresponding absorption spectra show the time scales relevant for spectra calculations.}
\end{figure}

\clearpage
\section{\label{sec:asymmetry}Asymmetry between S$_2\leftarrow$ S$_0$ absorption and S$_2\rightarrow$ S$_0$ emission spectra}
\begin{figure}[ht]
\includegraphics{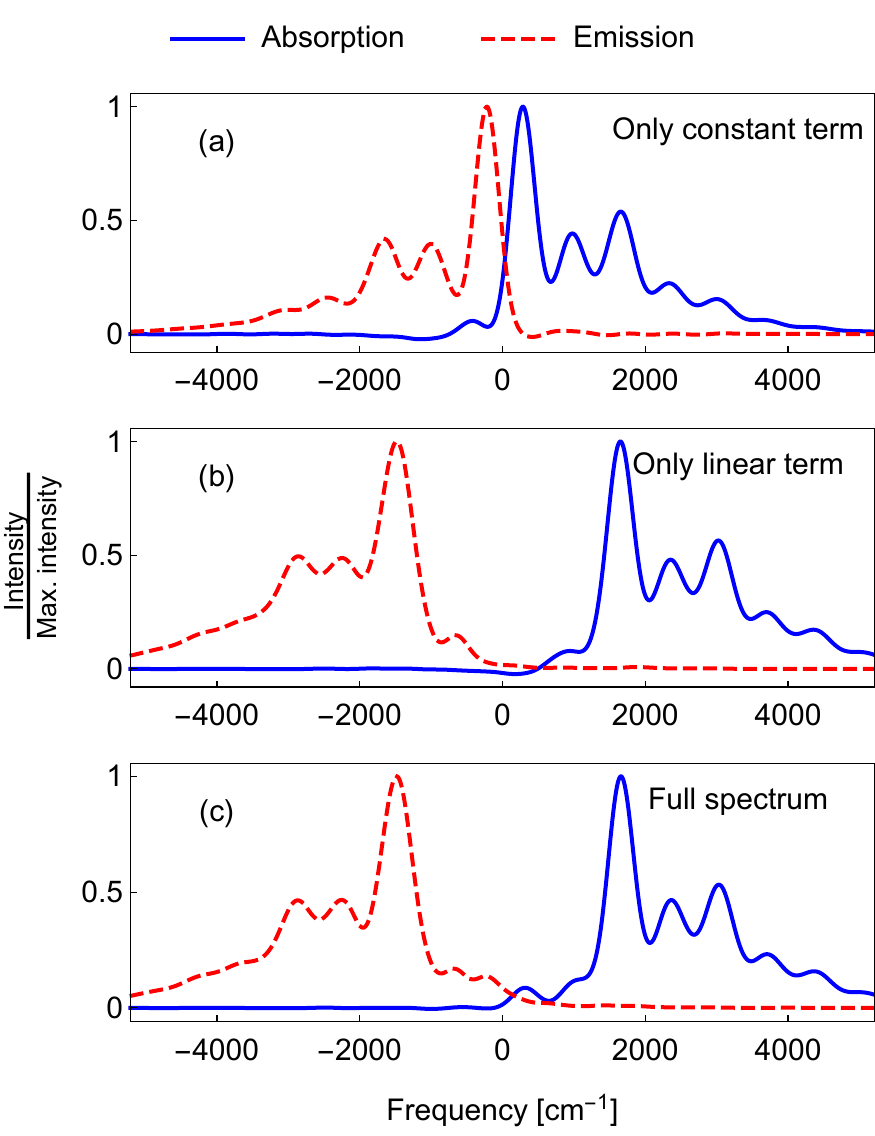}
\caption{\label{fig:MirrorImageS2_withoutDuschinsky}Vibrationally resolved S$_2\leftarrow$ S$_0$ absorption and S$_2\rightarrow$ S$_0$ emission spectra of azulene computed with (a) only Condon (constant) term, (b) only Herzberg--Teller (linear) term, (c) both Condon and Herzberg--Teller terms in the expansion of the transition dipole moment. Same as Fig.~5 of the main text, but neglecting the Duschinsky rotation between the ground- and excited-state normal mode coordinates, which is accomplished by setting the off-diagonal elements of the reference (adiabatic) Hessian (expressed in the initial-state normal mode coordinates) to zero.}
\end{figure}

\begin{figure}[ht]
\includegraphics{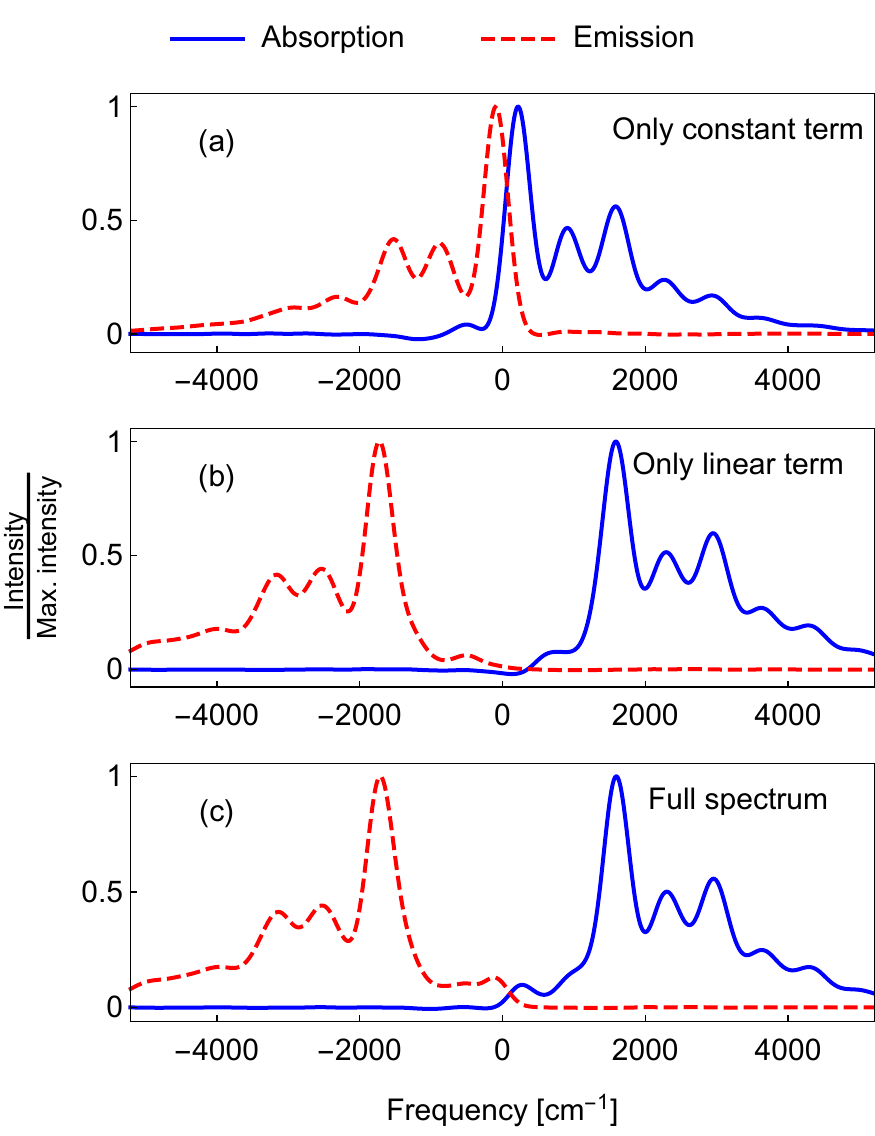}
\caption{\label{fig:MirrorImageS2_withoutDuschinskyPartial}Same as Fig.~\ref{fig:MirrorImageS2_withoutDuschinsky} but here neglecting only the coupling between the Herzberg--Teller active modes (listed in Table~\ref{tab:tdm}) in the simulations of the absorption spectra; the emission spectra are the same as in Fig.~5 of the main text.}
\end{figure}

\begin{table}[H]
\caption{Derivatives of the transition dipole moment (in atomic units) of
the azulene S$_{2} \leftarrow$ S$_{0}$ electronic transition with respect to the normal
mode coordinates. Only the largest terms of the transition dipole moment gradient are shown, derivatives with respect to other normal modes are either negligible or zero. The derivatives of the $z$ component of the transition dipole moment are all zero (the molecule lies in the $xy$ plane).}
\label{tab:tdm}\centering
\begin{tabular}{ccrrr}
\hline\hline
Frequency / cm$^{-1}$ & & $\partial \mu_{x} / \partial q \times 10^{2}$ & & $\partial \mu_{y} / \partial q \times 10^{2}$\\ \hline
 1784 &  &  $-0.86\quad$ & & $ 0.00\quad$ \\
 1660 &  &  $-2.50\quad$ & & $-5.31\quad$ \\
 1646 &  &  $-1.51\quad$ & & $ 0.69\quad$ \\
 1584 &  &  $ 0.00\quad$ & & $ 0.90\quad$ \\
 1514 &  &  $ 1.28\quad$ & & $-0.65\quad$ \\
 1490 &  &  $ 1.08\quad$ & & $ 2.32\quad$ \\
 1409 &  &  $-0.55\quad$ & & $ 0.00\quad$ \\
 1235 &  &  $-0.60\quad$ & & $-1.29\quad$ \\
 1074 &  &  $ 0.00\quad$ & & $ 1.05\quad$ \\
 \hline
 \hline
\end{tabular}%
\end{table}

\pagebreak

\bibliography{biblio45,additions_Azulene}